\documentstyle[aps,preprint,floats,epsfig]{revtex}
\input epsf

\def\simlt{\stackrel{<}{{}_\sim}}

\def\btabl{\begin{table}}   \def\etabl{\end{table}}
\def\bea{\begin{eqnarray}}   \def\eea{\end{eqnarray}}
\def\bnn{\begin{eqnarray*}}   \def\enn{\end{eqnarray*}}
\def\beq{\begin{equation}}   \def\eeq{\end{equation}}  
\def\btabu{\begin{tabular}}   \def\etabu{\end{tabular}}
\def\bec{\begin{displaymath}} \def\eec{\end{displaymath}}

\def\eqref#1{(\ref{#1})}
\renewcommand{\baselinestretch}{1.2}
\begin{document}
\newcommand{\bfig}{\begin{center}\begin{picture}}
\newcommand{\efig}[1]{\end{picture}\\{\small #1}\end{center}}
\newcommand{\flin}[2]{\ArrowLine(#1)(#2)}
\newcommand{\wlin}[2]{\DashLine(#1)(#2){3}}
\newcommand{\zlin}[2]{\DashLine(#1)(#2){5}}
\newcommand{\glin}[3]{\Photon(#1)(#2){2}{#3}}
\newcommand{\lin}[2]{\Line(#1)(#2)}
\newcommand{\sof}{\SetOffset}
\draft
\preprint{\vbox{\baselineskip=15pt
\rightline{CERN-TH/99-241}
\rightline{LPT Orsay-99/62}
\rightline{hep-ph/9908352}}}
\title{Constraints on a general 3-generation neutrino
mass matrix from neutrino data: application to the MSSM with R-parity violation}
\author{
A. Abada \footnote{
e-mail: abada@mail.cern.ch. On leave of absence from
the Laboratoire de Physique Th\'eorique, b\^at 210, Universit\'e de Paris XI,
91405 Orsay, France.},
M. Losada \footnote{
e-mail: losada@mail.cern.ch. On leave of absence from
the Universidad Antonio Nari\~{n}o, Santa Fe de Bogot\'a, Colombia.}
}
\vskip -0.5cm
\vspace{-0.5cm}
\address{{\small{\em 
Theory Division, CERN, CH-1211 Geneva 23, Switzerland}}}

\vskip 3cm 

\maketitle \begin{abstract}  
We consider a general symmetric $(3\times 3)$ mass matrix for three generations
of neutrinos. Imposing the constraints, from the atmospheric neutrino  
and solar neutrino anomalies as well as from the CHOOZ experiment, on
the mass squared differences and on the mixing angles, we identify
the ranges of allowed inputs for the 6 matrix elements. 
We apply our results to  Majorana left-handed neutrino masses generated at
tree level and through fermion--sfermion 
loop diagrams in the MSSM with
R-parity  violation. The present experimental results on neutrinos from
laboratories, cosmology and astrophysics are implemented to either put 
 bounds on trilinear ($\lambda_{ijk}, \lambda'_{ijk}$) and bilinear ($\mu_{e,\mu,\tau}$)
  R-parity-violating couplings 
or  constrain
combinations of products of these couplings.

Pacs numbers:  13.15.+g, 14.70.Bh,  95.30.Cq.
\end{abstract}

\vspace{0.5cm}
 
\leftline{CERN-TH/99-241}
\pacs{}
\renewcommand{\baselinestretch}{1.4}

\newpage \section{Introduction}
The current  experimental evidence strongly
suggests the existence  of a non-trivial structure 
of the mass matrix for 3 generations of neutrinos.
Indeed, deficits of the solar electron neutrino and atmospheric muon neutrino 
fluxes compared to
their theoretical predictions are strong evidence in favour of neutrino
oscillations as an explanation of these deficits. Furthermore, oscillations of neutrinos imply 
masses for neutrinos
and mixing angles, which relate the flavour basis to the mass eigenstate basis, as  is the case in the quark sector. 

Observations in favour of the oscillation solutions came recently with the 
Super-Kamiokande  collaboration \cite{ska} results, which have to be added to those of  other
 atmospheric neutrino  experiments (IMB \cite{imb}, Soudan \cite{soudan},
Kamiokande \cite{ka}) and solar neutrino experiments (Homestake \cite{davis}, Gallex 
\cite{gallex}, SAGE \cite{sage}, Kamiokande \cite{ks}, 
Super-Kamiokande \cite{sks}, MACRO 
 \cite{macro} and LSND 
\cite{lsnd}). The oscillation explanation
\footnote{Other possible solutions, which could accommodate the experimental
 neutrino
data, can involve flavour-violating interactions, 
or other more exotic possibilities such 
as the violation of Lorentz invariance \cite{coleman-glashow} or the
equivalence principle \cite{halprin}. However, in the case of atmospheric
neutrinos, it has been shown \cite{pakvasa}, \cite{fogli2} that neutrino decays or 
modification of special or general 
relativity do not fit the data.} of the solar neutrino problem, the atmospheric
neutrino anomaly and the LSND results suggest three very different values of 
 neutrino mass squared
differences, namely $\Delta m^2_{{\mathrm{sun}}} \ll\Delta m^2_{{\mathrm{atm}}}
 \ll\Delta m^2_{{\mathrm{LSND}}}$,  with $\Delta m^2_{{\mathrm{sun}}} \simlt 10^{-4}$
 eV$^2$, 
 $\Delta m^2_{{\mathrm{atm}}}\in [10^{-3},10^{-2}]$ eV$^2$ and 
 $\Delta m^2_{{\mathrm{LSND}}}\in [0.3,10]$ eV$^2$.  The evidence for
 $\bar\nu_\mu\to\bar\nu_e$ observed by LSND has not been confirmed or excluded
 by the KARMEN \cite{karmen} experiment. The MiniBooNE at FNAL or MINOS long-baseline 
 experiments
 could provide  the answer. 
Given the atmospheric and solar neutrino observations, and  excluding the  LSND results
 \footnote{In a forthcoming analysis, we
will include LSND results, which calls for a fourth neutrino species or a sterile
neutrino.},  
 the deduced pattern of neutrino masses and mixing requires
  two different neutrino mass squared differences 
 and three mixing angles.  Recent interest in this subject has produced a very large number
  of possible models, which can accommodate the data, 
 see refs.\cite{neutrinos} for more
details.
 
Our approach differs from most of the analyses in that we do not specify the model initially.
In a general manner, we will use the current relevant information on the mixing angles
 and mass squared differences to
constrain the  neutrino mass matrix elements in the flavour basis $(\nu_e,\nu_{\mu},\nu_{\tau})$. Consequently this  will constrain the
  parameters of the  involved model. The results of this work can be applied to any class of
  model with three neutrino generations. We also present an application
of our results of the specific Minimal Supersymmetric Standard Model (MSSM ) with R-parity violation.
We will also try to elucidate if any of
the different possible neutrino spectra (degenerate neutrinos, hierarchical, pseudo-Dirac 
\cite{barbieri}) are preferred for each solution of the solar anomaly combined
with SuperKamiokande and CHOOZ constraints. Even barring the presence of a sterile neutrino,
other interesting bounds  on neutrino masses arise
from cosmological or astrophysical sources, such as having a neutrino component of
hot dark matter (HDM). 
In addition, low energy experiments can also constrain neutrino masses;
 the strongest constraint,
 which is applicable  when the neutrinos are
 Majorana particles, arises from neutrinoless double beta decay. 
 We will also analyse our results in view of  the two latter bounds.

 The neutrino is massless in the 
Standard Model and, if massive, requires the presence of new physics.
For reviews, concerning models for neutrino masses and neutrino properties,  
see \cite{roulet,kayser,bilenky,valle} and references therein. 
If the neutrino has a Majorana mass, the
model must violate lepton number conservation. 
In supersymmetric (SUSY) extensions, gauge invariance and 
 renormalizability no longer ensure
 lepton number $L$ (or baryon number $B$) conservation. 
 The generalization of the MSSM, which allows  R-parity (${\mathrm{R}}_{\mathrm{P}}$) 
 violation
(${\mathrm{R}}_{\mathrm{P}}=(-1)^{L+3B+2S}$, where $L$, $B$, $S$ are the 
  lepton and 
 baryon number and the spin of the particle, respectively),  allows left-handed 
 neutrinos to obtain a
 Majorana mass, at tree level through mixing with the neutralinos, 
 and through loop diagrams that violate lepton number (in two units).
 We apply our general results to the MSSM with R-parity violation
 at one-loop order, allowing the presence of both bilinear and trilinear 
 ${\mathrm{R}}_{\mathrm{P}}$-violating couplings without imposing any hierarchy between
  the parameters.  We 
  thus constrain all  relevant ${\mathrm{R}}_{\mathrm{P}}$-conserving and non-conserving  couplings by imposing the 
constraints from atmospheric and solar neutrino anomalies and the CHOOZ experiment.
As a by-product, we see that in this model the constraints from neutrinoless double beta
decay and cosmology are {\it automatically}  satisfied.

The paper is organized as follows. In section II we expand on the inputs
used. In section III we establish our notation
for the general case of a 3-generation neutrino mass matrix. In section IV
we introduce our notation for an  application of our general results to
the MSSM with ${\mathrm{R}}_{\mathrm{P}}$-violation. We also give the results for
the tree-level neutrino mass matrix and the loop corrections to each matrix element.
  We start  by parametrizing 
the general neutrino eigenmasses
 and  eigenstates and use the neutrino data to 
 constrain the ranges of variation of the general inputs. In section IV.A we introduce a 
toy-model that corresponds to several limiting cases of the MSSM without R-parity conservation.
In section V we present our results for the general case and for the specific
case of Majorana masses arising in the R-parity violating model. 
Finally in section VI we summarize and
conclude.

\section{General inputs}
 In this work, we compute the masses and mixing angles of the neutrino
spectrum
  assuming three generations of neutrinos (this means that we are not taking 
   the LSND result into account). We then apply all relevant existing bounds on the neutrino masses
 and mixing angles to constrain the elements of our neutrino mass matrix in
the flavour basis.
 These inputs are from laboratories such as the CHOOZ experiment \cite{chooz},
  the relevant 
 mass squared differences
 $\Delta m^2$ and  mixing angles $\sin^2 2 \theta$ assuming 
  the oscillation solution as an explanation of the solar and
 atmospheric neutrino deficits.
  We then  check whether the 
 upper bound coming from neutrinoless beta decay
 $(\beta\beta)_{0\nu}$ \cite{{G-G},{M-H},{M-H-2}} 
 and also the global cosmological upper bound on the sum of the
 neutrino masses \cite{cosmo} are satisfied.
The best limit on the effective mass $m_{\mathrm{eff}}$ appearing in  $(\beta\beta)_{0\nu}$,
  and defined by
\bea\label{beta-0}
|m_{\mathrm{eff}}|=
 |\sum_i m_{\nu_i} U_{e i}^2|\le \sum_k  m_{\nu_k}| U_{e k}^2|,
\eea
 has been derived in the Heidelberg--Moscow $^{76}$Ge experiment \cite{G-G}: 
 $|m_{\mathrm{eff}}|\ < \  (0.5 \ -\  1.0)$ eV (at $90\% $ C.L.).
The CP phases that might appear in the mixing matrix $U_{e k}$ are not relevant
 in our derivation, because we only use the second r.h.s. term in 
eq. (\ref{beta-0}).
 Another direct upper bound comes from the measurement of the high energy part
of the $^3 {\mathrm{H}}\ \beta$-decay spectrum: the upper limits on the 
electron neutrino mass obtained in
the TROITSK \cite{troitsk} and MAINZ \cite{mainz} experiments are 
$m_\nu < 2.5$ eV and  $m_\nu <
3.4$ eV, respectively.
 However, these experiments suffer from some ambiguities referred
 to as ``the
negative mass squared problem", which is still not completely  
understood and we will therefore not use their bounds.

 There are also upper bounds on the neutrino masses coming from astrophysical
 and cosmological considerations such as the one from hot dark matter (HDM),
  which suggests that 
 $\sum_i m_{\nu_i} = m_{1} + m_{2} + m_{3}< $ few eV 
 \footnote{A recent analysis (observation of
 distant objects favouring a large cosmological constant instead of HDM)
  suggesting a non-zero
cosmological 
constant \cite{cosmological} shows that the bound of a few eV for the sum of the
neutrino masses is no longer imposed, since a 
 hot dark matter component is not a necessary ingredient in this case.}.

 Global fits to the data show that neutrino oscillations among three neutrino flavours are
sufficient to accommodate the solar and atmospheric data \cite{barbieri}, \cite{Fogli},
 \cite{Barger}, 
\cite{concha}, \cite{bahcall}. 
 Three different types of solar experiments (using different detectors:
  chlorine, gallium and Cherenkov)
 are sensitive to different solar neutrino energy ranges; to accommodate them,
  three regions of
 oscillation parameter space are allowed \cite{Fogli}, \cite{bahcall}, \cite{hata},
  which correspond to the 
 vacuum
 oscillation solution, MSW with a large mixing angle (MSW-LMA) 
 and MSW with a small mixing angle (MSW-SMA).
The 
estimates required to fit the data for the neutrino mass squared differences
 and the mixing angles are shown in table I.
 It is important to keep in mind that the neutrino oscillation scenario
can only restrict mass squared differences but not fix the absolute neutrino mass scale; 
moreover, it cannot distinguish whether the smallest mass splitting, which corresponds to the 
solar data, is between the two lightest mass eigenstates or the two heavier ones.

To fix our notation, for this case with three generations, 
the flavour states $\nu_l$  are expressed in terms of the  mass 
eigenstates $\nu_i$  using the $3\times3$
 mixing matrix $U $
\bea
\nu_\ell= \sum_{i=1}^3 U^*_{\ell i} \nu_i.
\eea
 We  hierarchically order  the mass eigenvalues and denote them  $m_i, \ i=1,2,3$, such that 
$$m_1\leq m_2\leq m_3.$$ 
In view of the ranges of the mass squared in table I, 
if the neutrino spectrum is hierarchical, then  $\Delta m_{12}^2=m_2^2-m_1^2$ is 
the parameter relevant to the solar case
and $\Delta m_{13}^2=m_3^2-m_1^2$ is the dominant one for the atmospheric case with 
$\Delta m_{12}^2\ll\Delta m_{13}^2$ and $\Delta m_{23}^2\sim\Delta m_{13}^2$. 
The corresponding mixing angles (see the parametrization below) are  $\sin^2
2\theta_{12}$ and $\sin^2
2\theta_{23}$, respectively. Besides the hierarchical mass spectrum, 
there exist two other alternative neutrino spectra  \cite{barbieri}, 
\cite{petcovs}, one in which
the three neutrinos are  quasi-degenerate  and another (called pseudo-Dirac) 
in which only two neutrinos
are quasi-degenerate and the third is much lighter (or heavier). 
In table II
  we show
the different possible regimes and indicate which is the corresponding  mass
 squared difference.

\begin{table}[hbt]
\begin{tabular}{|c|c|c|}
\hline
Experiment & $\Delta m^2$ (eV$^2$) & $\sin^2 2\theta$\\\hline
Atmospheric & $(1-8)\times 10^{-3}$& $0.85-1$\\\hline
Solar&&\\
 --MSW-{LMA}&$(3-30)\times 10^{-5}$ & $0.6\ -\ 1$\\
 --MSW-{SMA}&$(0.4-1)\times 10^{-5}$ & $10^{-3}\ -\ 10^{-2}$\\
 --Vacuum&$(0.5-8)\times 10^{-10}$ & $0.5\ -\ 1$\\\hline\hline
CHOOZ & $> 3 \times 10^{-3}$& $\sin \theta_{13}<0.22$ \protect\label{fitss}
\end{tabular}
\caption
{MSW-LMA, MSW-SMA and Vacuum stand for MSW large mixing angle, small mixing and vacuum 
oscillation solutions, 
respectively.}
\end{table}

\begin{table}[hbt]
\begin{tabular}{|c|c|c|}
Spectrum  &Solar& Atmospheric \\ \hline
Hierarchy  &$\Delta m_{12}^2$& $\Delta m_{13}^{2}$ \\ \hline
Degenerate & $\Delta m_{23}^2 \quad{\mathrm{or}}\quad \Delta m_{12}^{2}$ & 
$\Delta m_{13}^{2}$ \\ \hline
Pseudo-Dirac & $\Delta m_{23}^2 \quad{\mathrm{or}}\quad \Delta m_{12}^{2}$ & 
$\Delta m_{13}^{2}$ \protect\label{spectra}
\end{tabular}

\caption{Different possible regimes and corresponding  mass squared
difference. }
\end{table}

  \section{Computation of the mass and mixing matrix}

\subsection{Generic mass matrix}

We  consider a general $3\times 3$ symmetric mass matrix in the flavour basis
for three generations of neutrinos $(\nu_{e},\nu_{\mu},\nu_{\tau})$:

\bea\label{gen}{\cal{M}}=
 \left(
 \begin{array}{ccc}
 M_{11}& M_{12}& M_{13} \\
 M_{12}& M_{22}& M_{23} \\ 
 M_{13}& M_{23}& M_{33}
 \end{array}
 \right)\ .
\label{gralM}
  \eea

We construct the mass squared differences of the eigenvalues of ${\cal M}$
  and the mixing matrix, which relates the flavour basis to the mass eigenstates,
in a general way in terms of these input elements. We denote
the three mass eigenvalues by $m_{i}$ for $i=1,2,3$, which satisfies, as mentioned above, the
hierarchy $m_{1} \leq m_{2}\leq m_{3}$. For any given model, these masses and also the mixing matrix 
would be given in terms of the parameters of the model. 
We then apply the constraints, given in table I, 
individually and
combined. The combined constraints are obtained by requiring the atmospheric, CHOOZ and
one of the solar solutions to be satisfied simultaneously. In this way we can determine 
the region, in the
($\Delta m^2 \ ,  \sin^2 2\theta$) plane, that satisifies each of the three possible 
combinations of  constraints
simultaneously. The answer is then translated into terms of ranges where the 
parameters of the model vary so as  to fulfil neutrino data constraints.

\subsection{Generic mixing matrix}
Computing the mass eigenstates of the mass matrix (\ref{gen}), we obtain the 
 rotation matrix, which contains the relevant mixing angles. We use the Chau and 
 Keung parametrization of a $3\times
 3$ matrix \cite{C-K} given by

 \bea\label{mix}
U =  \left(
 \begin{array}{ccc}
 c_{12}\ c_{13}&s_{12}\ c_{13}& s_{13} e^{-i \delta}\\
 -s_{12}\ c_{23}-c_{12}\ s_{23}\ s_{13}e^{i \delta}
 &c_{12}\ c_{23}-s_{12}\ s_{23}\
 s_{13}e^{i \delta}&s_{23}\ c_{13}\\
 s_{12}\ s_{23}-c_{12}\ c_{23}\ s_{13}e^{i \delta}
 &-c_{12}\ s_{23}-s_{12}\ c_{23}\ s_{13}e^{i \delta}&
 c_{23}\ c_{13}
 \end{array}
 \right)\  {\mathrm{diag}}\left\{e^{i \alpha_1},e^{i \alpha_2} ,1\right\},
 \eea
where $c_{ij}=\cos (\theta_{ij})$ and  $s_{ij}=\sin (\theta_{ij})$, $\delta$ is the
Dirac CP phase and $\alpha_{1,2}$ are the Majorana ones 
(we have two additional CP phases in the case of Majorana particles).
 As previously mentioned,  we are going to work with a real mixing matrix and 
 will ignore CP phases because we are interested in our
 derivation in the modulus of the matrix elements, as we will see.
 
  The survival 
 probability $P_{\nu_e \to\nu_e}$, relevant to the case of solar fluxes, depends
 only on the first row of the mixing matrix in eq. (\ref{mix}), i.e. on 
 $|U_{ei}|^2$, with $i=1,2,3$. In the atmospheric case, the oscillation probability
 depends on the last column of (\ref{mix}), i.e. on 
 $|U_{\ell 3}|^2$, with $\ell=e,\mu,\tau$.
 The other elements of the matrix are not constrained by any direct experimental
 observation.
With this parametrization, we  directly obtain  the  parameters $\sin^2 2 \theta_{12}$ and 
$\sin^2 2 \theta_{23}$ and also the relevant CHOOZ parameter $\sin\theta_{13}$,
  which we need for our general analysis.

\section{Application: MSSM with R-parity violation}
 
The  specific ${\mathrm{R}}_{\mathrm{P}}$-violating model to which we apply our general 
results  has been 
extensively analysed  in the context of
pure phenomenology, as well as in trying to fit the data of neutrino anomalies. In refs. 
 \cite{barbieri}
\cite{Godbole},
\cite{haber},
 \cite{nardi}, \cite{banks}, \cite{nowakowski}, 
 \cite{Borzumati}, \cite{Hempfling}, \cite{white}, \cite{smirnov}, \cite{nilles} 
 different phenomenological aspects of this theory
have been analysed  with
respect to neutrino masses. In particular,  refs. \cite{nardi,banks} have established
in a basis-invariant way
that neutrino masses are always generated in these models even when universality of
the soft SUSY-breaking terms is assumed at some high scale.
Previous studies have also tried to constrain the different
 ${\mathrm{R}}_{\mathrm{P}}$-violating couplings that 
 appear in the MSSM Lagrangian,  considering only the effect of bilinear terms  
\cite{kaplan}--\cite{romao} 
or only of trilinear couplings \cite{choi}--\cite{bhattachar},  
 or from  both 
\cite{chunetal}--\cite{datta},
  solar and atmospheric neutrino data.
Both tree-level and one-loop effects have been considered.

As stated in the introduction here we do not bias our selection of 
${\mathrm{R}}_{\mathrm{P}}$-violating
 couplings, which would be contrary to a common practice of keeping only
one R-parity-violating parameter at a time. We also include all possible bilinear and trilinear parameters for the MSSM without
R-parity. In addition we do not impose any hierarchy between the trilinear couplings.
Finally, our results present the allowed ranges for these parameters using  the solar and
atmospheric anomalies and the result from the CHOOZ experiment only. 
We will present  our comparison  to previous analyses in the last section.

The most general renormalizable superpotential for the supersymmetric
 Standard Model with lepton-number violation is

\bea
W = \epsilon_{ab}[\mu_{\alpha} \hat{L}_{\alpha}^{a} \hat{H}_{u}^{b}} + \lambda_{\alpha\beta k}\hat{L_{\alpha}^{a} \hat{L}_{\beta}^{b}\hat{E}_{k}^{C}  + h_{ik}^{u} \hat{Q}_{i}^{a}
\hat{H}_{u}^{b} \hat{U}_{k}^{C}
 + \lambda'_{\alpha i k} \hat{L}_{\alpha}^{b} \hat{Q}_{i}^{a} \hat{D}_{k}^{C}],
\label{W}
\eea
the $(i,j,k)$ are flavour indices, $(a,b)$ are $SU(2)$ indices, and the $(\alpha,\beta)$ are
flavour indices running from 0 to 3. The $\hat{L}_\alpha$ are the doublet superfields
with hypercharge $Y=-1$. Note that the $\lambda$ couplings are antisymmetric in the first two
indices. The usual R-parity-preserving Lagrangian is obtained when only 
$\mu_{o}, \lambda_{0ik} = h_{ik}^{d}, \lambda'_{i0k}= h_{ik}^{d}$ are non-zero 
and we can identify $\hat{L}_{o} \equiv \hat{H}_{d}$.
In the model of eq. (\ref{W}) we have 9 additional $\lambda$ couplings and 27 new $\lambda'$ 
couplings compared
to the R-parity-conserving case. 
Note that thanks to the additional degrees of freedom, we can rotate in the flavour space of
the ``down-type'' scalar fields to set the vacuum expectation values 
of the sleptons to be zero \footnote{This can be done order by order 
in the loop expansion when one appropriately defines the mass matrices of
the Higgs sector \cite{Kong}.}. Henceforth, we will work consistently in this basis. 

We use the following
parametrization for 
 the  tree-level mass matrix in the basis of charged-lepton eigenstates: 
 
 \bea{\cal{M}}^{tree}_{\nu}=
 \left(
 \begin{array}{ccc}
 \alpha_e^2& \alpha_\mu\alpha_e& \alpha_e\alpha_\tau \\
\alpha_\mu\alpha_e& \alpha_\mu^2& \alpha_\mu\alpha_\tau \\
 \alpha_e\alpha_\tau& \alpha_\mu\alpha_\tau& \alpha_\tau^2
  \end{array}
 \right)\ ;
\label{Malphas}
  \eea
the corresponding expressions for the $\alpha_{i}$'s are given in appendix \ref{appendixa2}. 
This tree-level
mass matrix has only {\bf one} non-zero eigenvalue. It is thus necessary to
include additional corrections, which could provide  two different mass splittings.

The loop corrections, shown in fig. \ref{fig1}, from slepton--lepton and squark--quark
 loops to each element
of the mass matrix are given by \cite{haber},

\bea
(m_{qm})_{\mathrm{loop}}\! =\! {1\over 16 \pi^2} \left(\sum_{k,p} \lambda_{qkp}\lambda_{mpk}
m_{\ell}^{(k)} \sin 2\phi_{\ell}^{(p)}
\ln{M_{1}^{(p)}\over M_{2}^{(p)}} + 3 \sum_{k,p} \lambda'_{qkp}
\lambda'_{mpk} m_{q}^{(k)} \sin 2\phi_{q}^{(p)}
\ln{M_{q1}^{(p)}\over M_{q2}^{(p)}}\right).
\label{loopm}
\eea
Here the angles $\phi_{\ell}^{(p)}$ and $\phi_{q}^{(p)}$ are given in terms of the 
slepton and squark mass eigenstates for a given flavour $(p)$
$M_{1,2}$ and $M_{q1,q2}$ as
\bea
\sin 2\phi_{\ell}^{(p)}
\ln{M_{1}^{(p)}\over M_{2}^{(p)}} = m_\ell^{(p)} 
\left({2 X\over M^{(p)\ 2}_1-M^{(p)\ 2}_2} Ln
\left({M^{(p)}_1\over M^{(p)}_2}\right)\right)=
m_\ell^{(p)}\left({ X\over M^{(p)\ 2}_2} f(x_\ell^{(p)})\right)
\\
\sin 2\phi_{q}^{(p)}
\ln{M_{q1}^{(p)}\over M_{q2}^{(p)}} = m_q^{(p)} \left({2 X\over
M^{(p)\ 2}_{q1}-M^{(p)\ 2}_{q2}} Ln
\left({M^{(p)}_{q1}\over M^{(p)}_{q2}}\right)\right)=m_q^{(p)}\left({ X\over M^{(p)\
2}_{q2}} f(x_q^{(p)})\right),
\eea
where $m^{(p)}_{\ell,q}$ are lepton and  quark masses of flavour $p$;
 $X=A +\mu \tan\beta$   \cite{Kane}, 
which we take to be universal, and provides a mixing term between the 
left- and right-handed squarks and sleptons. The function $f$ is defined to be
\bea
f(x)=-{\ln \ x\over 1-x},\quad
x_\ell^{(p)}= \left({M_1^{(p)}\over M_2^{(p)}}\right)^2\quad {\mathrm{and}}\quad
x_q^{(p)}= \left({M_{q1}^{(p)}\over M_{q2}^{(p)}}\right)^2.
\eea
We consider here that we are in the down-quark mass eigenstate basis, and that the
$\lambda,\lambda'$ couplings have been redefined in terms of the couplings
appearing in the superpotential and of the corresponding Cabibbo--Kobayashi--Maskawa
matrix elements; however, for simplicity we do not introduce additional notation.

In this model, in order to simplify, we drop the slepton-flavour dependence from 
$ { X\over M^{(p)\ 2}_2} f(x_\ell^{(p)})$ and ${ X\over M^{(p)\
2}_{q2}} f(x_q^{(p)})$ and consider them to be universal in the slepton and
squark sector respectively.
 We are  then led to

\bea
(m_{qm})_{\mathrm{loop}} = {1\over 16 \pi^2} X\left({f(x_\ell) \over M^{2}_2} 
\sum_{k,p} \lambda_{qkp}\lambda_{mpk} m_{\ell}^{(k)}m_{\ell}^{(p)} + 3{f(x_q) \over M_{q2}^2}
 \sum_{k,p} \lambda'_{qkp}\lambda'_{mpk} m_{q}^{(k)} m_{q}^{(p)}\right).
\label{loopmnew}
\eea
We use the following
parametrization for 
 the  one-loop mass matrix written in the same  basis as the tree-level mass matrix: 
 \bea{\cal{M}}^{\mathrm{loop}}_{\nu}=
 \left(
 \begin{array}{ccc}
 m_{11}& m_{12}& m_{13} \\
 m_{12}& m_{22}& m_{23} \\ 
 m_{13}& m_{23}& m_{33}
 \end{array}
 \right)\ .
  \eea
 The loop mass matrix elements $m_{ij}$ are given in appendix \ref{appendixa3},
  where we have kept only
 the relevant contributions after employing the mass hierarchy selection from the  charged-lepton and
 down-quark sector.
  
  Finally, the total mass matrix we consider is 
  \bea
  {\cal{M}_{\nu}}={\cal{M}}^{\mathrm{tree}}_{\nu}
+{\cal{M}}^{\mathrm{loop}}_{\nu}.
\eea

An important remark is that there are additional loop corrections  that can generate 
neutrino masses through  the 
sneutrino mass splitting in these models \footnote{We especially
thank S. Davidson for discussions on this point.}. An additional gaugino--sneutrino (Higgs)
loop appears where the R-parity violation is now contained in
the mass-mixing term between the sneutrino and antisneutrino.  
Naively, it can be
shown  that  this  effect is  small when  the soft mass vector 
$B = B_{\mu_{\alpha}}$ and the
 vacuum expectation value vector of the down-type scalar fields $v_{\alpha}$  are aligned 
\cite{davidson-king,grossman2}. However, a full treatment of this
effect  requires a one-loop calculation of the scalar mass matrices (Higgses and
sleptons), which is beyond the scope of this paper.

\begin{figure}[hbt]
\hspace{2.5cm}\epsfxsize=6cm\epsfbox{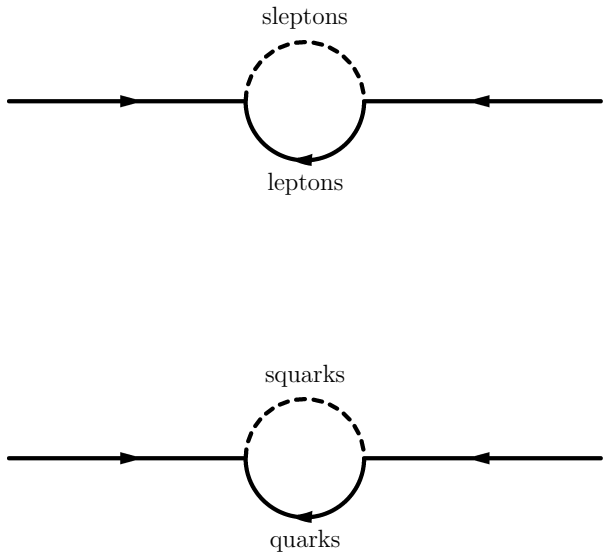}
\vspace*{-2.5cm}\vskip -4cm
\hspace{2.5cm}\caption[]{\it{One-loop Feynman diagrams contributing to the neutrino masses.}} 
 \protect\label{fig1}
\end{figure}

 \subsection{Toy model}
 In order to have an idea of how to constrain the whole space of parameters given 
 by the different  $\lambda$ and $\lambda'$ involved in the loop contribution to the
 mass and the $ \mu_{e,\mu,\tau}$ appearing in the tree-level mass with the neutrino
 data, we put ourselves in the naive picture where all the $\lambda$  couplings are
 equal and also the $\lambda'$ couplings. 
 In fact we are naively assuming in eq. (\ref{loopmnew})
 (see appendix \ref{appendixa3}) that 
 $\lambda_{133}=\lambda_{233}=\lambda_{333}$ and 
  $\lambda'_{133}=\lambda'_{233}=\lambda'_{333}$.\\
   We thus have 
   two universal trilinear R-parity
 couplings, $\lambda$ and $\lambda'$, and three bilinear couplings,
  $\mu_e$, $\mu_{\mu}$ and $\mu_\tau$.
 In this toy model, our mass matrix can be written  as follows
  \bea{\cal{M}_{\nu}}=
 \left(
 \begin{array}{ccc}
 K_{1}+K_{2}+\alpha_e^2&   K_{1}+K_{2} +\alpha_e \alpha_{\mu}& K_{2}+ \alpha_e\alpha_\tau\\
 K_{1}+K_{2}+\alpha_e \alpha_{\mu}&  K_{1}+K_{2}+\alpha_{\mu}^2&  K_{2}+
 \alpha_\mu\alpha_\tau \\ 
K_{2}+ \alpha_e\alpha_\tau &K_{2}+ \alpha_\mu\alpha_\tau  & K_{2}+\alpha_\tau^2
 \end{array}
 \right),
\label{MRp}
 \eea
where 

\begin{eqnarray}
K_{1} & =& {X\over 16 \pi^2} {f(x_\ell) \over M^{2}_2} \left(
 \lambda^2 m_{\tau}^2  \right)\ , \nonumber \\
K_{2}&=&3  {X\over 16 \pi^2} {f(x_q) \over M^{q2}_2} \left( \lambda'^2 m_{b}^2 \right)\ .
\label{AAB}
\end{eqnarray}
 Generically this matrix has three non-zero eigenmasses.
Under several different approximations (i.e. $K_1=0$, or $K_2=0$, or $\alpha_{\mu}= \alpha_{e}$),
this mass matrix yields only {\it two} non-zero
eigenvalues

 \bea
 U^{t}\times{\cal{M}} \times U={\mathrm{diag}}\{0,m_2, m_3\}.
 \eea
The different physical motivations  (or limits) for these approximations correspond to the cases where:
\begin{itemize}
\item $\lambda=0$ or $f(x_\ell) = f(x_{q})$ and $M_{2} = M_{2}^{q}$ gives $K_1=0$ due to the hierarchy $m_{b} \gg m_{\tau}$.
\item $\lambda'=0$ or, if we take the limit in which the squarks decouple,  $K_2=0$.
\item assume some hierarchy between the bilinear terms in the Lagrangian, such that $\alpha_{e}\sim \alpha_{\mu}
\ll \alpha_\tau$.

\end{itemize}

\section{Results}
\subsection{General results}
\begin{figure}[hbt]

\vspace{-10pt}

\centerline{\hspace{-3.3mm}
\epsfxsize=8cm\epsfbox{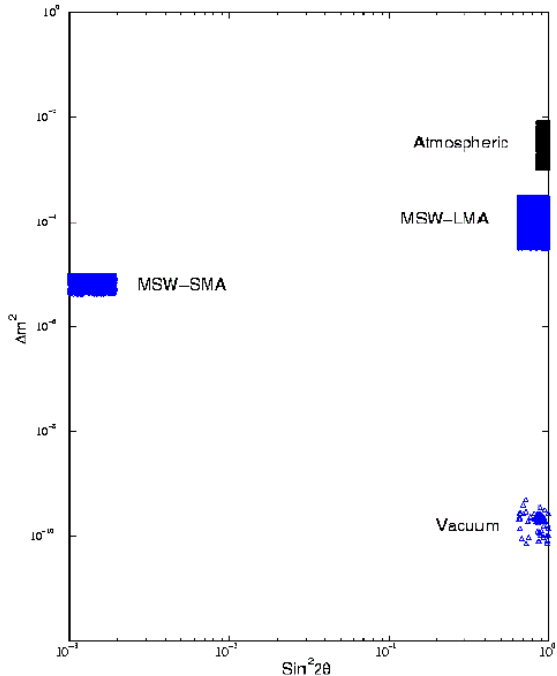}}
\caption[]{{
$\Delta m^2$ in units of eV$^2$ as a function of $\sin^2 2\theta$. 
The atmospheric and different solar solutions that are
presented result from a  combined  fit with CHOOZ and SuperK 
constraints.\\
The atmospheric solution is represented by the parameters 
($\Delta m^2_{13}$, $\sin^2 2\theta_{23}$).   The 
MSW-LMA and  MSW-SMA solar
solutions are represented  by  the parameters ($\Delta m^2_{12}$ or $\Delta m^2_{23}$ and $\sin^2
2\theta_{12}$) 
while the vacuum solution is represented  by the parameters 
($\Delta m^2_{12}$, $\sin^2 2\theta_{12}$).}} 
\protect\label{all-general-50}
\end{figure}

\vskip 0.5cm
We have performed a general scan of parameter space made up by 
the six matrix elements that appear in eq. (\ref{gralM}). In this section
we will simply present the results, and  our discussion
of them is given in the next section.
We present in fig. \ref{all-general-50}
the allowed points  on the ($\Delta m^2$, $\sin^2 2\theta$) plane, which
satisfy the combined constraints (CHOOZ, SuperK and solar data) 
for the  possible solutions
of the solar and atmospheric anomalies.
Allowing only combined solutions of solar, SuperK and CHOOZ data 
restricts the available region of  parameter space. 
We also see in  fig. \ref{all-general-50}  
that   all three possible solutions for the solar anomaly, CHOOZ and 
atmospheric SuperK bounds can be  simultaneously satisfied.
  After several runs, we choose the interval of
variation 
 of the matrix elements  $M_{ij}$ in such a way that, combined to CHOOZ, SuperK 
 constraints and 
  one (or  all) of the solar solutions is (are) possible. 
   For example, in order to have a combined fit of the vacuum solution for
solar anomaly, together
   with CHOOZ and SuperK constraints then the matrix elements have been
varied in the interval $ -0.5\ {\mathrm {eV}}\le M_{ij}\le 0.5 \ {\mathrm
{eV}}.$\\

We also note that the density of points (solutions) for
 Vacuum + CHOOZ + SuperK is smaller than the one for MSW-SMA + CHOOZ + SuperK
 which in turn is much smaller than the one corresponding to  MSW-LMA + CHOOZ + SuperK.

Tables III, IV and V, which are given in the next subsections,
  present the
 allowed ranges of the input parameters (matrix elements) that satisfy the
different solar oscillation solutions together with the CHOOZ {\it and}
the atmospheric SuperK constraints. 

To study the spectrum for each of the possible combined solutions we give
samples of the different mass spectra for different values of the matrix
elements $M_{ij}$. For illustration, we have present our plots for the
spectrum as a function of
 the matrix element $M_{13}$,
 which varies when we scan over parameter space. We also plot
 the effective mass relevant for neutrinoless double beta 
decay  and the sum of the mass eigenvalues. 
Our results for the latter two quantities show that for the specific range
that we have chosen to satisfy the combined fit of solar, CHOOZ and
atmospheric constraints, the present bounds are always satisfied. In fact,
for Majorana neutrinos,
 a slightly stronger bound on the effective mass can be placed by our
analysis and will be given below in each subsection. 
 This
bound is valid when we vary the mass matrix elements  in the  ranges that 
are specified below for each subcase.

\vfill
\subsubsection{MSW-LMA}

\begin{figure}[h]


\centerline{\hspace{-3.3mm}\epsfxsize=8cm\epsfbox{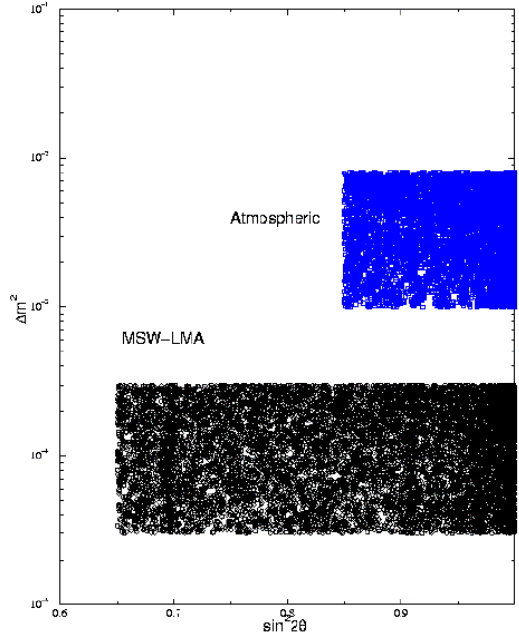}}
\caption[]{{{
$\Delta m^2$ in units of eV$^2$ as a function of $\sin^2 2\theta$. 
The atmospheric and solar solutions that are
presented result from a  combined  fit of MSW-LMA,  CHOOZ and SuperK 
constraints.\\
The atmospheric solutions are represented with the parameters 
($\Delta m^2_{13}$, $\sin^2 2\theta_{23}$) while  the 
MSW-LMA solar
solutions are presented  with the parameters ($\Delta m^2_{12}$ 
or $\Delta m^2_{23}$ and $\sin^2
2\theta_{12}$).}}}

\protect\label{LMACSK}
\end{figure}

\begin{figure}[hbt]


\centerline{\hspace{-3.3mm}
\epsfxsize=8cm\epsfbox{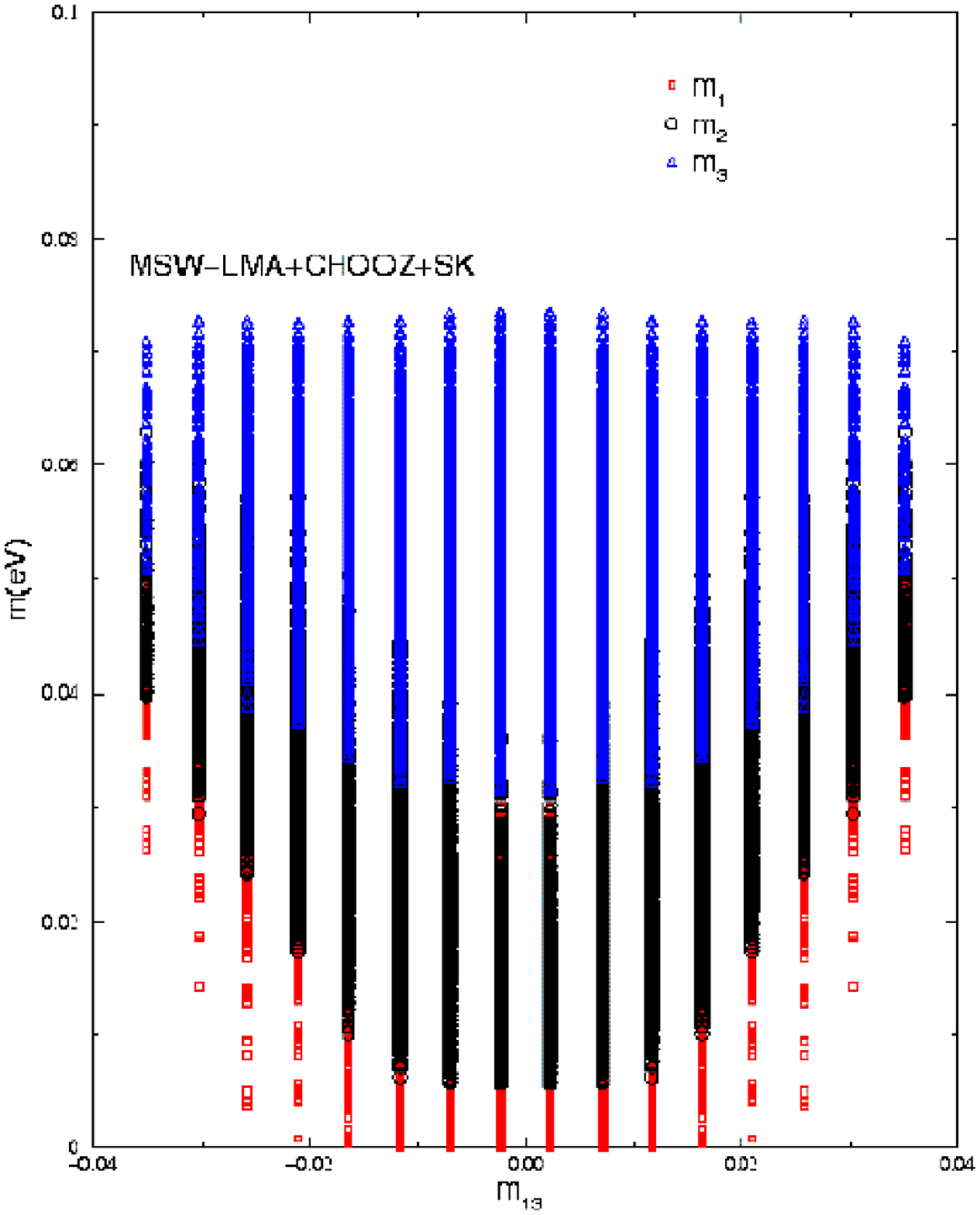}
\hspace{-0.1cm}
\epsfxsize=8cm\epsfbox{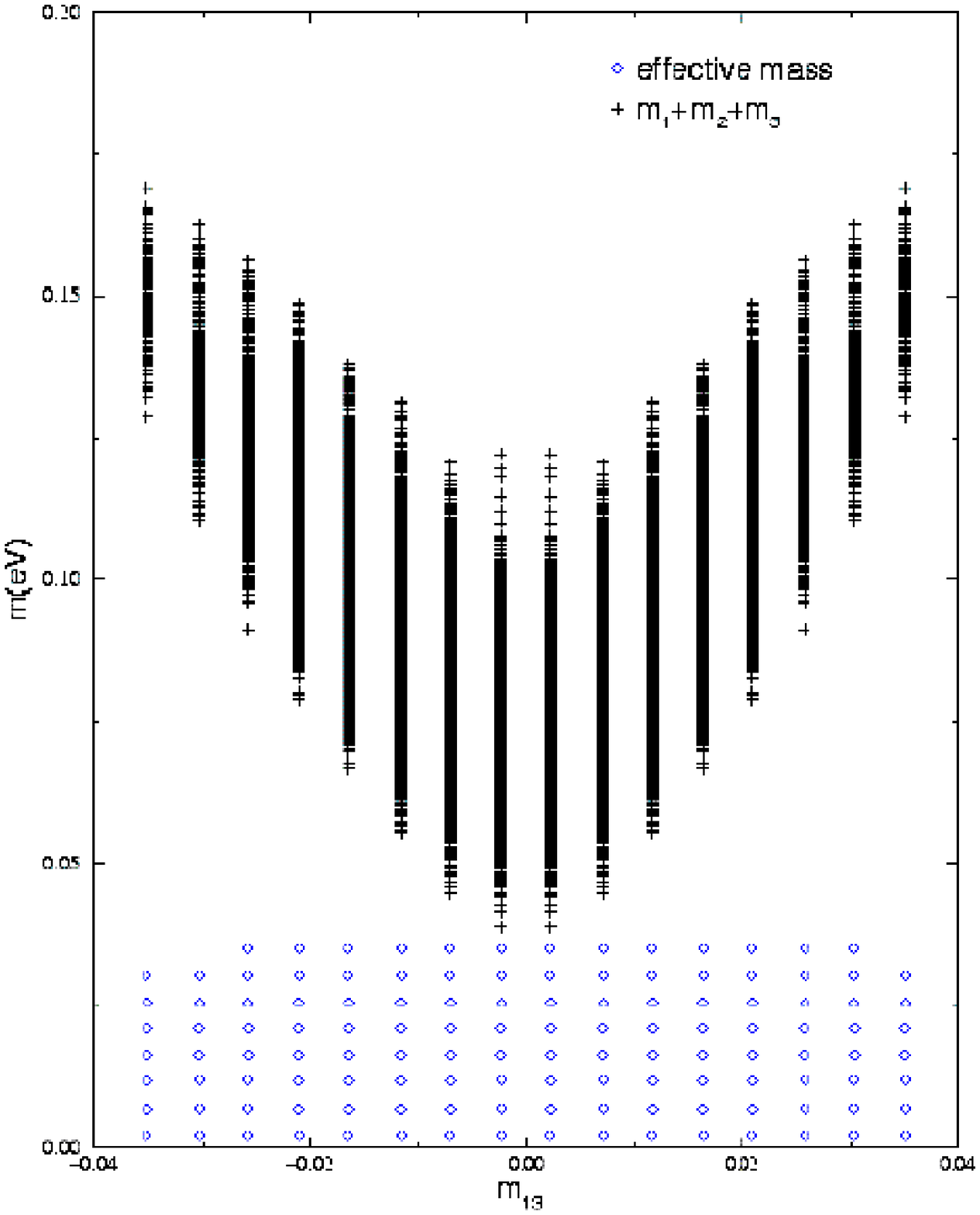}
}

\vspace*{-0.1cm}

\centerline{\hspace{1cm} (a) \hspace{7.5cm} (b)}

\vspace*{-0.1cm}
\caption{The spectrum 
presented results from a  combined  fit with CHOOZ and SuperK and MSW-LMA solar
constraints.
 Mass spectrum (a) which presents the three possible mass eigenvalues 
(eV).
  The effective mass and the sum of eigenmasses are given in (b).}
\protect\label{spect-LMA}
\end{figure}
\begin{table}[hbt]
\vspace{-10pt}
\begin{tabular}{|c|c|c|}
Inputs & Minimum  & Maximum\\\hline
$M_{11}(10^{-2}{\mathrm eV})$ & $-2.7$  & $7.3$\\\hline
$M_{12}(10^{-2}{\mathrm eV})$ & $-4.3$& $4.3$\\\hline
$M_{13}(10^{-2}{\mathrm eV})$ & $-3.9$  & $3.9$\\\hline
$M_{22}(10^{-2}{\mathrm eV})$ & $-1.1$& $8.3$\\\hline
$M_{23}(10^{-2}{\mathrm eV})$ & $-7.4  $  & $7.4$\\\hline
$M_{33}(10^{-2}{\mathrm eV})$ & $-5.0$& $8.9$\protect\label{tab1}
\end{tabular}
\caption{Allowed ranges of the input parameters  
 that satisfy { MSW-LMA}, SuperK and CHOOZ constraints simultaneously.
They also fulfil neutrinoless double beta decay and cosmological constraints.
This illustration is for a model where the allowed inputs vary initially in a
given interval, i.e., 
$M_{ij}\in [-1.0,+1.0]$ eV .}
\end{table} 

For a given model, imposing MSW-LMA, atmospheric SuperK and CHOOZ constraints 
simultaneously determines the allowed 
ranges for the input parameters $M_{11}$ (eV), $M_{22}$ (eV),
  $M_{12}$ (eV), $M_{23}$ (eV), $M_{33}$ (eV), $M_{13}$ in (eV), 
 summarized in table III. In order to obtain these results we have initially allowed all six matrix
elements to vary in the interval  $\left[-1.0,+1.0\right]$ eV.

Figure \ref{LMACSK} shows the available regions in
parameter space from  the combined constraint of
SuperK data together with CHOOZ and the MSW-LMA solution.

It is clear from fig. \ref{LMACSK} that, with this scan of parameter space, we
can cover most of the allowed region that satisfies the combined constraints
mentioned above.

 In figs. \ref{spect-LMA}a and \ref{spect-LMA}b we plot the
  results of the individual mass eigenvalues
and the effective mass (only valid for Majorana masses) together with the sum of
the eigenmasses, which is the relevant quantity for HDM when we impose the
constraints simultaneously.
We can see from fig. \ref{spect-LMA}a that the hierarchical spectrum is
preferred for small values of our parameter  $M_{13}$.\\
Using the results presented in
 fig. \ref{spect-LMA}b, for the entire  range of variation of 
 the matrix element $M_{13}$, 
 a strong bound  can be placed  on the  neutrinoless double beta decay
effective mass, i.e. for $M_{13}\in [-4,+4] \times 10^{-2}$ eV 
(while the other matrix elements vary in the intervals given in table III), 
$m_{{\mathrm {eff}}}\simlt 5\times 10^{-2}$ eV.

\begin{table}[hbt]
\begin{tabular}{|c|c|c|}
Inputs & Minimum  & Maximum\\\hline
$M_{11}(10^{-2}{\mathrm eV})$ & $-4.0$  & $2.9$\\\hline
$M_{12}(10^{-2}{\mathrm eV})$ & $-4.0$& $4.0$\\\hline
$M_{13}(10^{-2}{\mathrm eV})$ & $-4.0$& $4.0$\\\hline
$M_{22}(10^{-2}{\mathrm eV})$ & $-1.3$& $8.3$\\\hline
$M_{23}(10^{-2}{\mathrm eV})$ & $-8.0 $  & $8.0$\\\hline
$M_{33}(10^{-2}{\mathrm eV})$ & $-4.0$& $5.9$\protect\label{tab2}
\end{tabular}
\caption{Allowed ranges of the input parameters that satisfy 
{ MSW-SMA}, SuperK and CHOOZ constraints simultaneously.
They also fulfil neutrinoless double beta decay and cosmological constraints. 
This illustration is for a model where the allowed inputs vary initially in 
the interval $[-1.0,+1.0]$ eV .}
\end{table}



\subsubsection{MSW-SMA}

Imposing MSW-SMA, SuperK and CHOOZ constraints simultaneously gives the 
ranges for the input parameters 
 summarized in table IV. As in the previous section, this illustrative table 
 is valid for a
 given model in which the mass matrix elements vary initially, before applying
 the solar, atmospheric and CHOOZ constraints,  in the range
 $\left[-1.0,+1.0\right]$ eV.

\begin{figure}[h]

\vspace{-20pt}

\centerline{\hspace{-3.3mm}
\epsfxsize=8cm\epsfbox{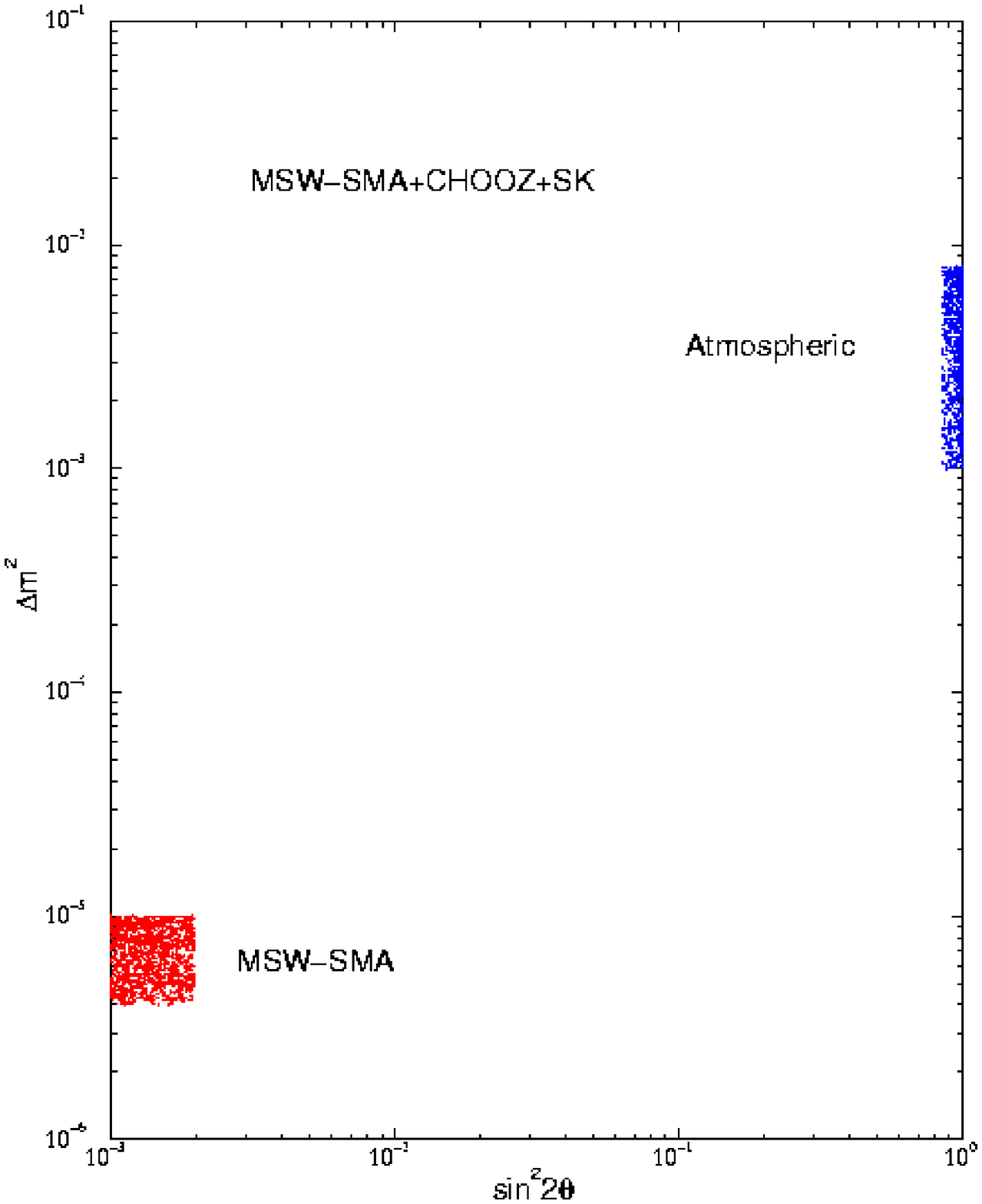}
\hspace{0.1cm}
\epsfxsize=8cm\epsfbox{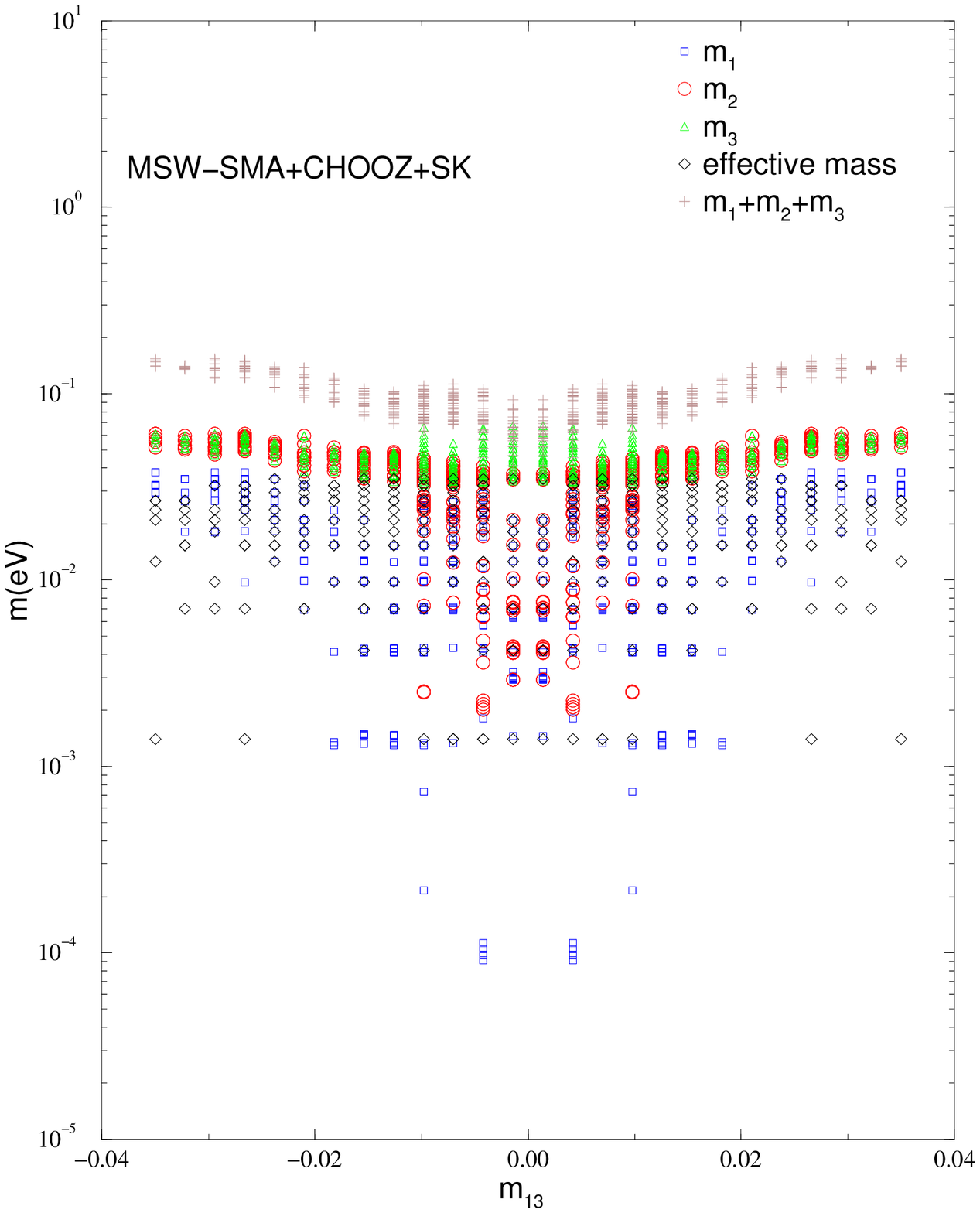}}


\centerline{\hspace{1cm} (a) \hspace{6.5cm} (b)}
\caption{In (a) we represent $\Delta m^2$ in units of eV$^2$ as a function 
of $\sin^2 2\theta$. 
The atmospheric solutions are represented with the parameters 
($\Delta m^2_{13}$, $\sin^2 2\theta_{23}$) and the 
MSW-SMA solar
solutions  with the parameters ($\Delta m^2_{12}$ 
or $\Delta m^2_{23}$ and $\sin^2
2\theta_{12}$) . In (b),
the mass spectrum is represented together with $m_{{\mathrm{eff}}}$
 and the sum of
the eigenvalues (eV). The atmospheric  and solar solutions that are
presented result from a  combined  fit with CHOOZ and SuperK 
constraints.}
\protect\label{SMACSK}
\end{figure}

Figure \ref{SMACSK}a gives the combined constraint coming from
SuperK data together with the  CHOOZ bound and the MSW-SMA solution.
 We have noticed  that with our scan of the parameter space
the resulting  density of points in fig. \ref{SMACSK}a is much smaller than 
 in the case of MSW-LMA (fig. \ref{LMACSK}).

We show in  fig. \ref{SMACSK}b the results of the
 individual mass eigenvalues, the effective mass
 and the values of $\Sigma m_{i}$ when we impose the
constraints.  Here
once again there is a  bound 
 $m_{{\mathrm {eff}}}\simlt few \times 10^{-2}$ eV for 
  $M_{13}\in [-4.0,+4.0]\times 10^{-2}$ eV 
  (the other matrix elements take values in the ranges given in table IV).

\newpage\subsubsection{Vacuum Solar Solution}


 In this case, there are  few solutions that satisfy 
simultaneously all of  the constraints.
One can see in fig.  \ref{V+V-C}a  
 how the allowed points on the  ($\Delta m^2$, $\sin^2 2\theta$) plane
are reduced compared to the  density of points in the cases of MSW-LMA 
(fig. \ref{LMACSK}) 
and MSW-SMA 
(fig. \ref{SMACSK}a).\\
The allowed
ranges for the input parameters,  
after  imposing the 
vacuum oscillation solution, SuperK and  CHOOZ constraints simultaneously, are 
 summarized in table V. This illustrative table 
 is valid for a
 given model in which the mass matrix elements vary initially, before applying
 the solar, atmospheric and CHOOZ constraints,  in the range 
 $\left[-0.5,+0.5\right]$ eV.\\
In fig. \ref{V+V-C}b,  where the spectrum of eigenmasses is represented
together with the effective mass $m_{{\mathrm{eff}}}$ and the sum over
 the eigenmasses, a salient feature is the presence of   
  two nearly degenerate eigenvalues ($m_1\sim m_2$).  
  Here we also have a strong bound
on the effective mass  $m_{{\mathrm {eff}}}\simlt {\mathrm{few}}10^{-2}$ eV 
as we vary 
$M_{13}$ in  $[-4.0,4.0]\times 10^{-2}$ eV 
(the other matrix elements take values in the ranges given in table V).



 \begin{figure}[ht]


\centerline{\hspace{-3.3mm}
\epsfxsize=8cm\epsfbox{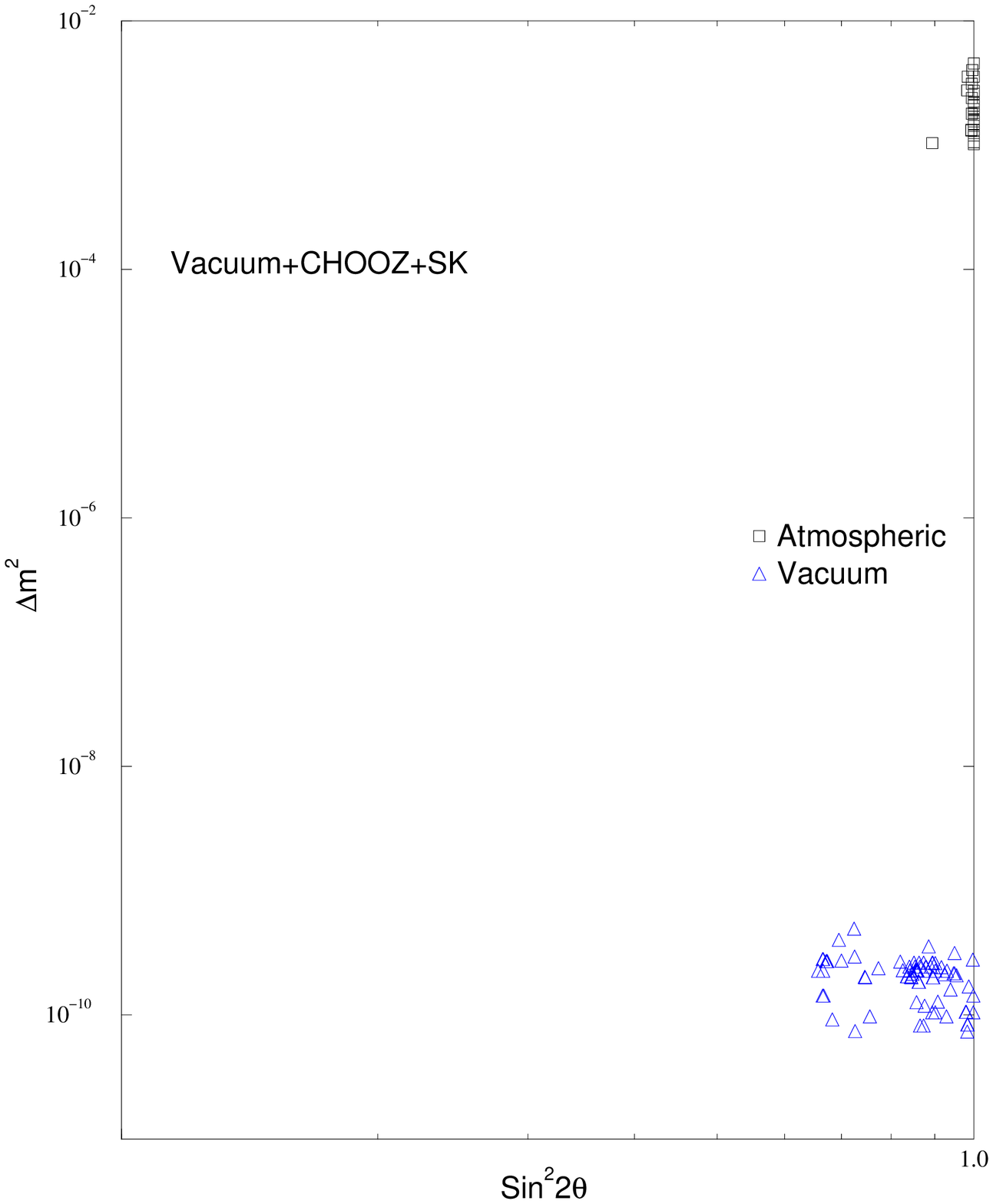}
\hspace{-0.1cm}
\epsfxsize=8cm\epsfbox{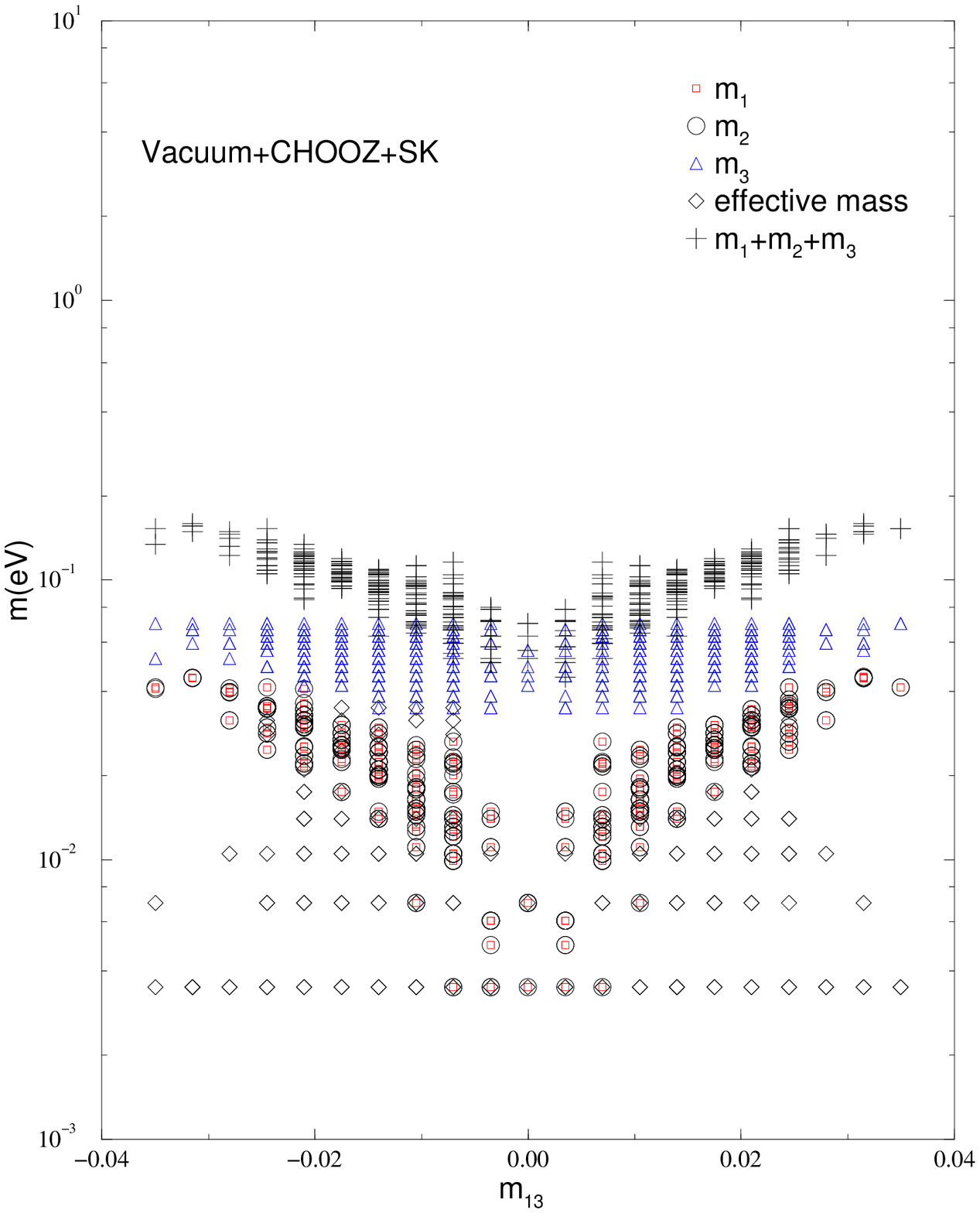}}


\centerline{\hspace{1cm} (a) \hspace{6.5cm} (b)}

\caption{ In (a) we represent $\Delta m^2$ in units of eV$^2$ 
as function of $\sin^2 2\theta$. The 
square and the triangle 
refer to the atmospheric solution ($\Delta m^2_{13}$, $\sin^2 2\theta_{23}$)
and the vacuum solar
solution ($\Delta m^2_{12}$, $\sin^2 2\theta_{12}$),  respectively. In (b),
the mass spectrum is represented together with $m_{{\mathrm{eff}}}$
 and the sum of
the eigenvalues (eV).}
\protect\label{V+V-C}
\end{figure}

\begin{table}[h]
\begin{tabular}{|c|c|c|}
Inputs & Minimum  & Maximum\\\hline
$M_{11}(10^{-2}{\mathrm {eV}})$ & $-0.7$  & $1.3$\\\hline
$M_{12}(10^{-2}{\mathrm {eV}})$ & $-4.0 $& $4.0$\\\hline
$M_{13}(10^{-2}{\mathrm {eV}})$ & $-4.0 $  & $4.0$\\\hline
$M_{22}(10^{-2}{\mathrm {eV}})$ & $+0.98$& $4.1$\\\hline
$M_{23}(10^{-2}{\mathrm {eV}})$ & $ -3.5 $  & $4.1$\\\hline
$M_{33}(10^{-2}{\mathrm {eV}})$ & $+0.98$& $4.1$\protect\label{tab3}
\end{tabular}
\caption{Allowed ranges of the input parameters that satisfy solar {Vacuum},  SuperK and 
CHOOZ constraints simultaneously.
They also fulfil neutrinoless double beta decay and cosmological constraints.
This illustration is for a model where the allowed inputs vary initially in 
the interval $[-0.5,+0.5]$ eV .}
\end{table}

 \clearpage
 \vfill
 \newpage
 \newpage\subsection{Bounds on R-parity-violating couplings from neutrino
 data}

Our approach to the numerical analysis is to use low energy input without any
reference to high-scale physics.

 The loop contributions are proportional to
different R-parity-violating parameters from those which control the tree-level terms.
 Three different possibilities can be considered:\\
- the tree-level contributions are much  larger than the loop contributions,\\
- the tree-level contributions are of the same order as the loop contributions,\\
- the tree-level contributions are much smaller than the loop contributions.

Applying the results of our general scan of parameter space, we are in 
fact including all of the above-mentioned cases in our analysis.
As each combination of constraints defines a different allowed
range of parameters we will present below the
corresponding bounds on R-parity-violating couplings  separately for
each case.

\subsubsection{MSW-LMA}

Imposing MSW-LMA, SuperK and CHOOZ constraints simultaneously defines the 
ranges for the input parameters $K_1$ ({{eV}}) (which is proportional to $\lambda^2)$, 
$K_2$ ({{eV}}) (which is proportional to $\lambda'^2)$, 
$\alpha_e$ ({{eV}}$^{1/2}$), $\alpha_\mu$ ({{eV}}$^{1/2}$), 
$\alpha_\tau$ ({{eV}}$^{1/2}$), which we have summarized in table VI.

\begin{table}[hbt]
\begin{tabular}{|c|c|c|}
Inputs & Minimum  & Maximum\\\hline
$K_1({\mathrm{eV}})$ & $0.0$  & $3.0\times 10^{-2}$\\\hline
$K_2({\mathrm{eV}})$ & $0.0$& $2.7\times 10^{-2}$\\\hline
$\alpha_e\ ({\mathrm{eV}}^{1/2}$)& $-0.114$& $0.154$\\\hline
$\alpha_\mu\ ({\mathrm{eV}}^{1/2}$)& $-0.280$& $0.28$\\\hline
$\alpha_\tau\ ({\mathrm{eV}}^{1/2}$)& $-0.238$& $0.238$\protect\label{tab1Rp}
\end{tabular}
\caption{Allowed ranges of input parameters
 that satisfy MSW-LMA, SuperK and CHOOZ constraints simultaneously.
They also fulfil neutrinoless beta decay and cosmological constraints.}
\end{table}

Figure \ref{LMACSKtoy} shows the regions that satisfy the combined constraint coming from
SuperK data together with CHOOZ and the MSW-LMA solution.

\begin{figure}[h]


\centerline{\hspace{-3.3mm}
\epsfxsize=8cm\epsfbox{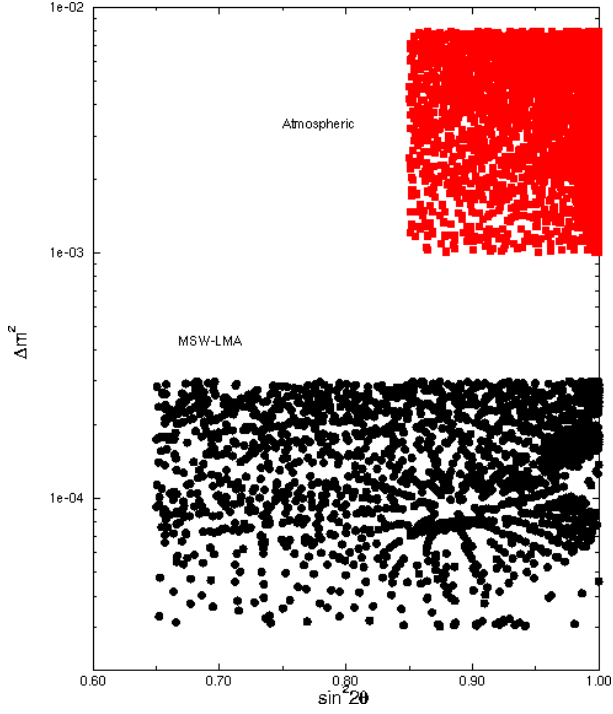}}
\vspace{10pt}
\caption{{
$\Delta m^2$ in units of eV$^2$ as a function of $\sin^2 2\theta$. The 
square and the circle 
refer to the atmospheric solution ($\Delta m^2_{13}$, $\sin^2 2\theta_{23}$)
and to the  MSW-LMA solar
solution ($\Delta m^2_{12}$, $\sin^2 2\theta_{12}$),  respectively.}}
\protect\label{LMACSKtoy}
\end{figure}

 We plot in fig.  \ref{LMAspec-toy}a the results of the individual 
 mass eigenvalues, and in fig. \ref{LMAspec-toy}b,  
  the effective mass $m_{{\mathrm{eff}}}$ together with the sum over the
eigenmasses  when we impose all the
constraints simultaneously. For illustration, we show how the mass spectrum 
and
the effective mass vary as a function of   $x=|\alpha_e({\mathrm{eV}}^{1/2})|$. One can read from fig. \ref{LMAspec-toy}b that in the range of parameters 
  $x\in[0,0.1]$,   we can put a strong bound on the effective mass, that is 
  $m_{{\mathrm{eff}}}\simlt 5\times 10^{-2}$ eV.
  We can also determine, in the same region of the parameter  space, a strong limit on
  $m_1+m_2+m_3 \simlt 0.2$ eV.
\vspace{0.2cm}
\begin{figure}[h]


\centerline{\hspace{-3.3mm}
\epsfxsize=8cm\epsfbox{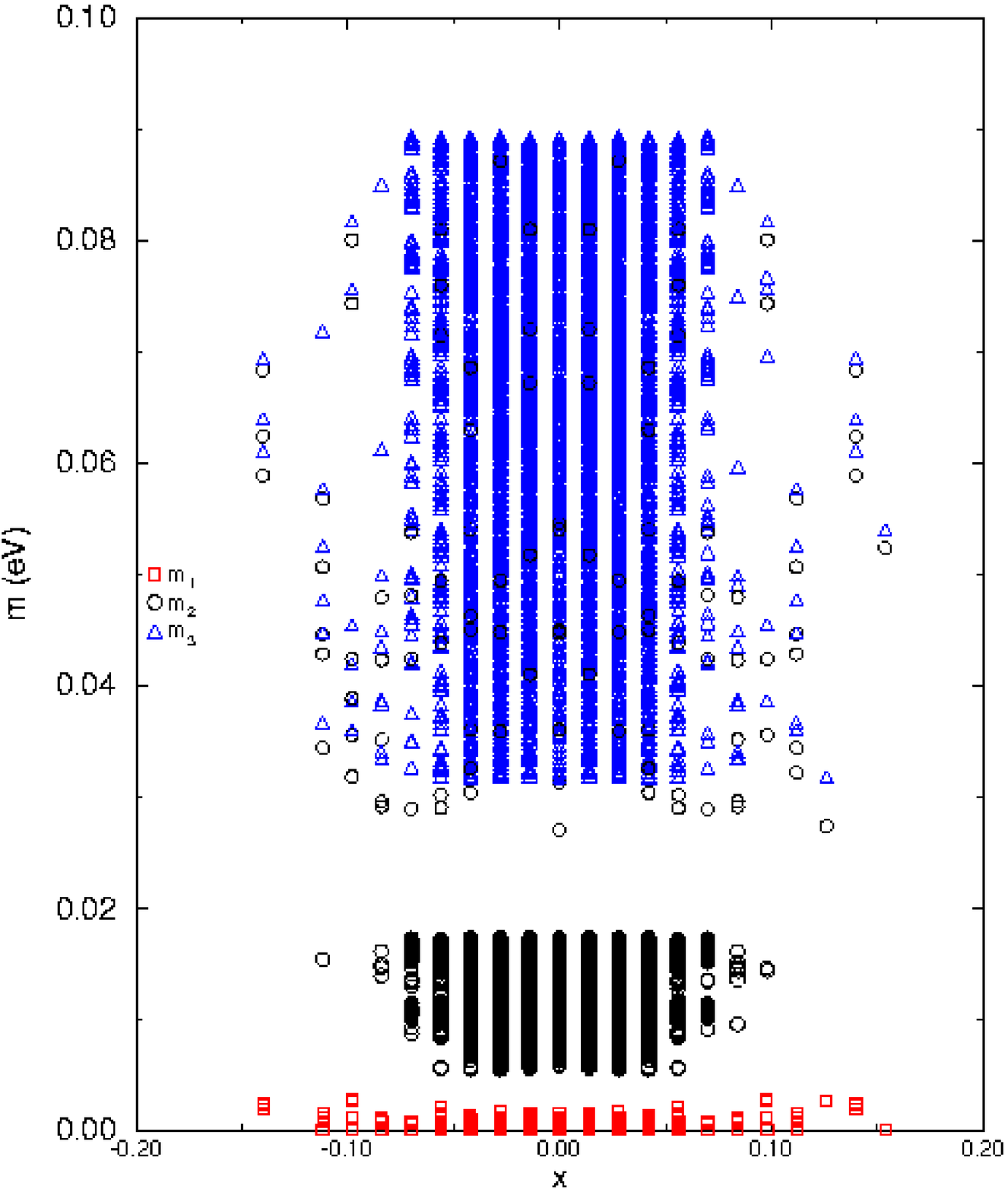}
\hspace{0cm}
\epsfxsize=8cm\epsfbox{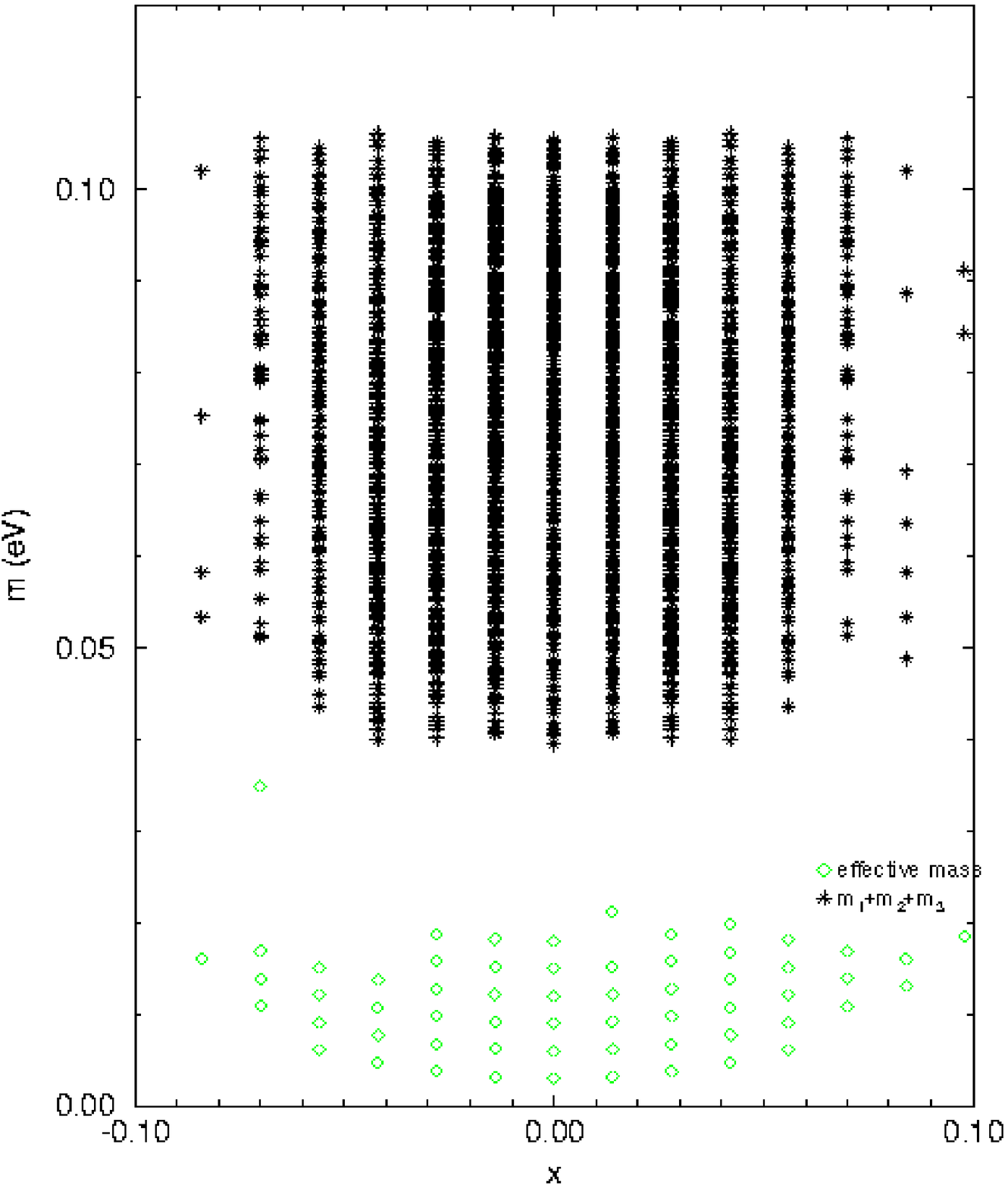}}


\centerline{\hspace{1cm} (a) \hspace{7.5cm} (b)}

\caption{The mass spectrum (eV) is represented in (a).
  The effective mass and the sum of eigenmasses (eV) are represented in (b). 
  This is when the MSW-LMA solution, the atmospheric one and CHOOZ constraint are imposed
  simultaneously.}
\protect\label{LMAspec-toy}
\end{figure}
 
 Applying our  results from table VI 
  to the specific mass matrix of eq. (\ref{MRp}) 
we find that the following bounds on the ${\mathrm {R}}_{\mathrm{P}}$-violating parameters exist
for each solar solution combined with SuperKamiokande and CHOOZ data constraints. For the 
trilinear couplings we have

\bea
\lambda^2 \leq {16 \pi^2 M_{2}^{2}\over X f(x_{l}) m_{\tau}^{2}}
 \times \left({\bf 3.0\times 10^{-2} }\right),
\label{lamba}
\eea

\bea
\lambda'^2 \leq {16 \pi^2 M_{2}^{q^{2}}\over 3 X f(x_{q}) m_{b}^{2}}
 \times 
\left({\bf 2.7\times 10^{-2}}\right).
\label{lamba'}
\eea

Using the results of the appendix \ref{appendixa2} we can also give a bound on the bilinear couplings,
\bea
\mu_{e}^{2} \leq {4 \det M \over m_{Z}^{2}\cos^{2}\beta (M_{1} +
 \tan^{2}\theta_{W}M_{2})} \times
\left({\bf 0.154} \ {\mathrm{eV}}^{-1/2}\right)^2 ,
\label{mui1}
\eea

\bea
\mu_{\mu}^{2} \leq {4 \det M \over m_{Z}^{2}\cos^{2}\beta (M_{1} +
 \tan^{2}\theta_{W}M_{2})} \times
\left({\bf 0.28} \ {\mathrm{eV}}^{-1/2}\right)^2 ,
\label{mui2}
\eea
\bea
\mu_{\tau}^{2} \leq {4 \det M \over m_{Z}^{2}\cos^{2}\beta (M_{1} +
 \tan^{2}\theta_{W}M_{2})} \times
\left({\bf 0.238} \ {\mathrm{eV}}^{-1/2}\right)^2 .
\label{mui3}
\eea

\subsubsection{MSW-SMA}
Imposing MSW-SMA, SuperK and CHOOZ constraints simultaneously determines the 
ranges for the input parameters  given in table VII.

\begin{table}[hbt]
\begin{tabular}{|c|c|c|}
Inputs &Minimum  & Maximum\\\hline
$K_1({\mathrm{eV}})$ & $0.0$  & $3.0\times 10^{-3}$\\\hline
$K_2({\mathrm{eV}})$ & $0.0$& $3.0\times 10^{-3}$\\\hline
$\alpha_e\ ({\mathrm{eV}}^{1/2}$)& $-4.2\times 10^{-2}$& $4.2\times 10^{-2}$\\\hline
$\alpha_\mu\ ({\mathrm{eV}}^{1/2}$)& $-0.28$& $0.28$\\\hline
$\alpha_\tau\ ({\mathrm{eV}}^{1/2}$)& $-0.21$& $0.21$\protect\label{tab2Rp}
\end{tabular}
\caption{Allowed ranges of the input parameters
 that satisfy MSW-SMA, SuperK and CHOOZ constraints simultaneously.
They also fulfil neutrinoless beta decay and cosmological constraints.}
\end{table}

Figure \ref{SMACSK-toy} gives the combined constraint coming from
SuperK data together with CHOOZ and the MSW-SMA solution. As mentioned previously for
 our analysis of a generic mass matrix, the density of 
 points is much smaller than in the case of 
 the MSW-LMA solution (fig. \ref{LMACSKtoy}).

\begin{figure}[h]


\centerline{\hspace{-3.3mm}
\epsfxsize=8cm\epsfbox{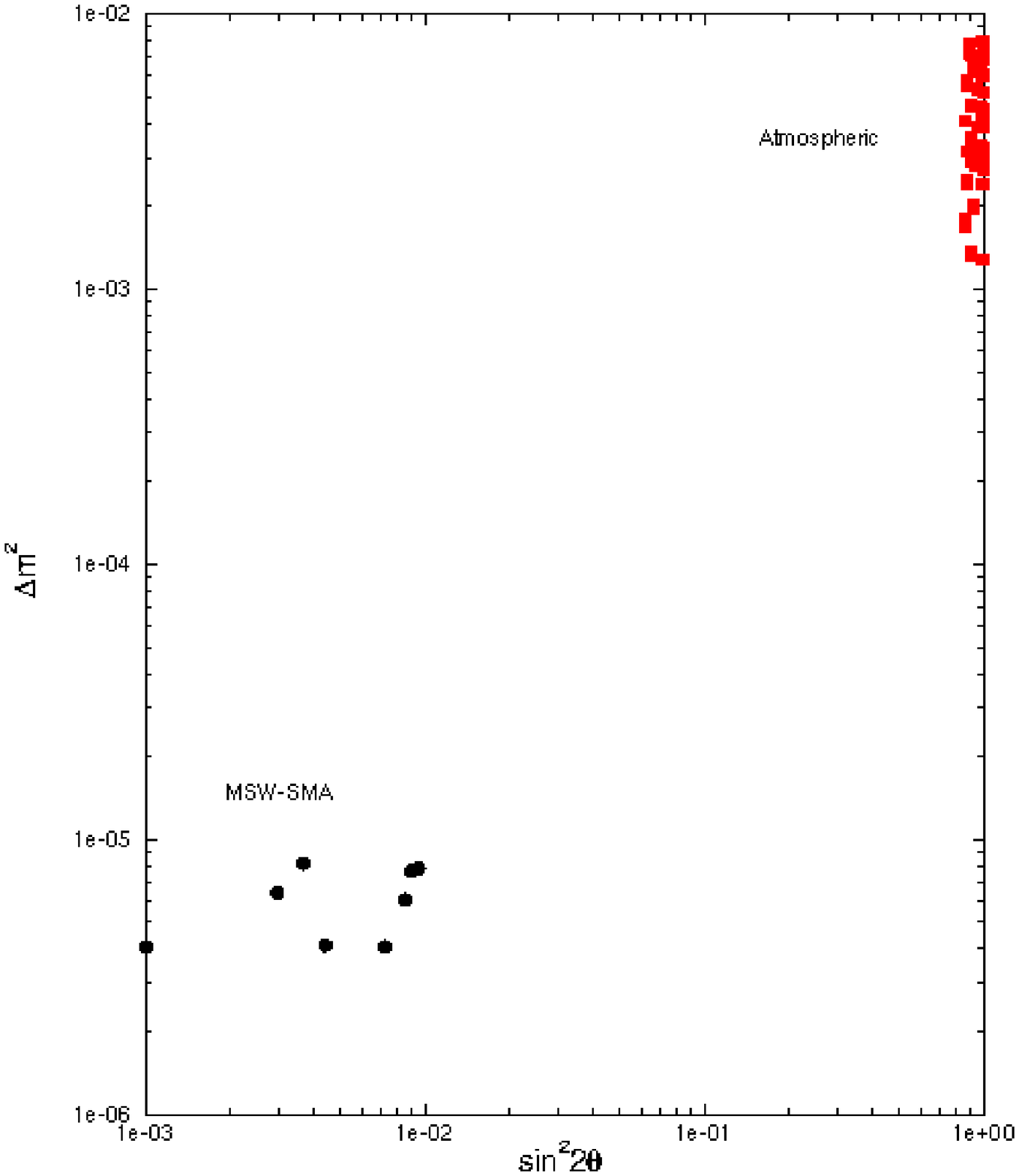}}

\caption{{
$\Delta m^2$ in units of eV$^2$ as a function of $\sin^2 2\theta$. The 
square and the circle 
refer to the atmospheric solution ($\Delta m^2_{13}$, $\sin^2 2\theta_{23}$)
and the MSW-SMA solar
solution  ($\Delta m^2_{12}$, $\sin^2 2\theta_{12}$),  respectively.}}
\protect\label{SMACSK-toy}
\end{figure}

 As in the previous subsection by applying our  results from table VII
 to the specific mass matrix of eq. (\ref{MRp}) 
we find  the following bounds on the ${\mathrm{R}}_{\mathrm{P}}$-violating parameters 
for each solar solution combined with SuperKamiokande and CHOOZ data constraints:

\bea
\lambda^2 \leq {16 \pi^2 M_{2}^{2}\over X f(x_{l}) m_{\tau}^{2}}\times
 \left({\bf 3.0\times 10^{-3}}\right)\ ,
\label{lambas}
\eea

\bea
\lambda'^2 \leq {16 \pi^2 M_{2}^{q^{2}}\over 3 X f(x_{q}) m_{b}^{2}}\times
 \left({\bf 3.0\times 10^{-3}}\right)\ ,
\label{lamba's}
\eea 

\bea
\mu_{e}^{2} \leq {4 \det M \over m_{Z}^{2}\cos^{2}\beta (M_{1} + 
\tan^{2}\theta_{W}M_{2})} \times\left({\bf 4.2\times 10^{-2} }\  
 {\mathrm{eV}}^{-1/2}\right)^2 \ ,
\label{mui1s}
\eea

\bea
\mu_{\mu}^{2} \leq {4 \det M \over m_{Z}^{2}\cos^{2}\beta (M_{1} + 
\tan^{2}\theta_{W}M_{2})} \times\left({\bf 0.28 }
 \ {\mathrm{eV}}^{-1/2}\right)^2 \ ,
\label{mui2s}
\eea

\bea
\mu_{\tau}^{2} \leq {4 \det M \over m_{Z}^{2}\cos^{2}\beta (M_{1} + 
\tan^{2}\theta_{W}M_{2})} \times\left({\bf 0.21 }
\ {\mathrm{eV}}^{-1/2}\right)^2 \ .
\label{mui3s}
\eea

\newpage
\subsubsection{Vacuum}

In our present application, we find that  
imposing vacuum oscillations, atmospheric and CHOOZ constraints simultaneously or 
 vacuum oscillation and CHOOZ  constraints simultaneously offers no 
solution of the solar and atmospheric anomalies. We have determined
which are the allowed  ranges for the input parameters by 
imposing only vacuum solution constraints; these are 
 summarized in table VIII. 

\begin{table}[hbt]
\begin{tabular}{|c|c|c|}
Inputs & Minimum  & Maximum\\\hline
$K_1({\mathrm{eV}})$ & $0.0$  & $6.0\times 10^{-3}$\\\hline
$K_2({\mathrm{eV}})$ & $0.0$& $1.20\times 10^{-2}$\\\hline
$\alpha_e\ ({\mathrm{eV}}^{1/2}$)& $-0.35$& $0.35$\\\hline
$\alpha_\mu\ ({\mathrm{eV}}^{1/2}$)& $-0.35$& $0.35$\\\hline
$\alpha_\tau\ ({\mathrm{eV}}^{1/2}$)& $-0.35$& $0.35$\protect\label{tab3Rp}
\end{tabular}
\caption{Allowed ranges for input parameters that satisfy Vacuum oscillation constraints.
They also fulfil neutrinoless beta decay and cosmological constraints.}
\end{table}

 We determine the  bounds on the ${\mathrm{R}}_{\mathrm{P}}$-violating parameters  from
the  results from table VIII. 
These are

\bea
\lambda^2 \leq {16 \pi^2 M_{2}^{2}\over X f(x_{l}) m_{\tau}^{2}} \times 
\left({\bf 6.0\times 10^{-3}}\right)
\label{lambav}
\eea

\bea
\lambda'^2 \leq {16 \pi^2 M_{2}^{q^{2}}\over 3 X f(x_{q}) m_{b}^{2}} \times
\left({\bf 1.20\times 10^{-2}}\right)
\label{lamba'v}
\eea 

\bea
\mu_{i}^{2} \leq {4 \det M \over m_{Z}^{2}\cos^{2}\beta (M_{1} + 
\tan^{2}\theta_{W}M_{2})}
\times \left( {\bf 0.35}  \ {\mathrm{eV}}^{-1/2}\right)^2 ,\quad i=e,\mu,\tau\ .
\label{muiv}
\eea

\newpage\subsubsection{Limits on $\lambda$, $\lambda'$ and $\mu_i$} 
Given our results for the bounds on the ${\mathrm{R}}_{\mathrm{P}}$-violating couplings, 
the analysis in terms of ${\mathrm{R}}_{\mathrm{P}}$-conserving SUSY parameters 
is straightforward. Our
general results can be analysed by considering the SUSY particle spectrum in
detail; we will present these results in a forthcoming paper \cite{abada-losada2}.
A common assumption used to place bounds in this model is to take  all 
${\mathrm{R}}_{\mathrm{P}}$-conserving
 mass parameters to be of the same order,
 $M_{susy}$. For the particular case where $M_{susy}=100$ GeV,
  $f(x) \rightarrow 1$ and $\tan\beta =2$, the bounds are 
\begin{itemize}
\item MSW-LMA + CHOOZ + SuperK
\bea
&&\lambda^2 \leq 1.5\times  10^{-7} \ , \hspace{1cm} 
\lambda'^2 \leq  7.02\times 10^{-9} \ ,\hspace{0.5cm} \nonumber\\
&&\mu_{e} \leq 1.9 \times 10^{-4}\  {\mathrm{GeV}},\hspace{0.5cm}
\mu_{\mu} \leq 3.4 \times  10^{-4}\  {\mathrm{GeV}},\hspace{0.5cm}
\mu_{\tau} \leq 2.9 \times 10^{-4}\  {\mathrm{GeV}}\ .
\label{num1l}
\eea
\item MSW-SMA + CHOOZ + SuperK
\bea
&&\lambda^2 \leq 1.5\times  10^{-8}\ , \hspace{1cm} 
\lambda'^2 \leq  7.8\times 10^{-10} \ ,\hspace{0.5cm} \nonumber\\
&&\mu_{e} \leq 5.2 \times 10^{-5}\  {\mathrm{GeV}},\hspace{0.5cm}
\mu_{\mu} \leq 3.4 \times  10^{-4}\  {\mathrm{GeV}},\hspace{0.5cm}
\mu_{\tau} \leq 2.6 \times 10^{-4}\  {\mathrm{GeV}}\ .
\label{num2s}
\eea
\item Vacuum
\bea
&&\lambda^2 \leq 3.0\times  10^{-8}\ , \hspace{1cm} 
\lambda'^2 \leq  3.1\times 10^{-9} \ ,\hspace{0.5cm} \nonumber\\
&&\mu_{i} \leq 4.3 \times 10^{-4}\  {\mathrm{GeV}},\quad i=e,\mu,\tau\  .
\label{num3v}
\eea
\end{itemize}

\subsection{Dependence of the $(\Delta m^{2},\sin^2 2\theta)$ solar 
and atmospheric parameters on the ${\mathrm{R}}_{\mathrm{P}}$-violating 
 parameters}

We now  plot in figs. \ref{LMA-x-sun}, \ref{LMA-x-atm} and \ref{LMA-sum}, 
 the different solar and
atmospheric parameters as functions of the 
${\mathrm{R}}_{\mathrm{P}}$-violating parameters $K_1
\propto \lambda^2$, $K_2
\propto \lambda'^2$, $\alpha_{e,\mu,\tau}$ and also
$\alpha^2=\alpha_e^2+\alpha_\mu^2+\alpha_\tau^2$. 

In figs. \ref{LMA-x-sun} and \ref{LMA-x-atm}, where 
the constraints from CHOOZ,
SuperK and the MSW-LMA solution are satisfied simultaneously, we have plotted the
solar parameters $\Delta m^2_{12}$ and $\sin^2 2\theta_{12}$
 (fig. \ref{LMA-x-sun})  and the atmospheric variables  
 $\Delta m^2_{13}$ and $\sin^2 2\theta_{23}$ (fig. \ref{LMA-x-atm}).
Figure \ref{LMA-sum} illustrates the dependence of the solar and
atmospheric parameters on the sum of the squares of the 
${\mathrm{R}}_{\mathrm{P}}$-violating bilinear terms for the combined fits
of SuperK, CHOOZ and MSW-LMA (figs. \ref{LMA-sum}a and \ref{LMA-sum}c) as well as for
Super, CHOOZ and MSW-SMA (figs. \ref{LMA-sum}b and \ref{LMA-sum}d).
 Here  one can observe how the region of parameter
space
  from 
the fit that satisfies MSW-LMA, CHOOZ and SuperK simultaneously is considerably larger than 
 the
available region where 
we  fit the  MSW-SMA, CHOOZ and SuperK atmospheric constraints.
\newpage
\begin{figure}[hbt]


\centerline{\hspace{-3.3mm}
\epsfxsize=8cm\epsfbox{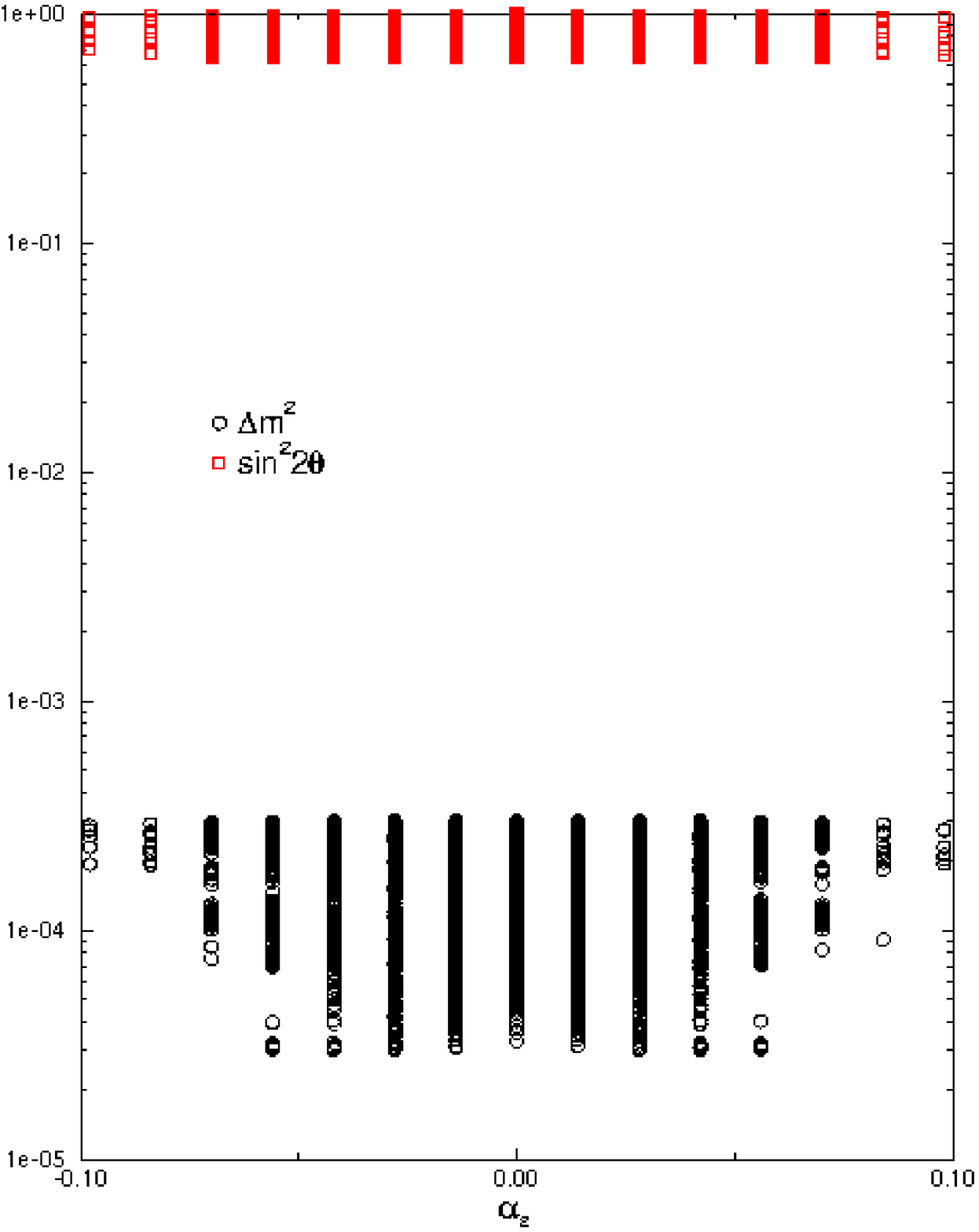}
\hspace{-0.1cm}
\epsfxsize=8cm\epsfbox{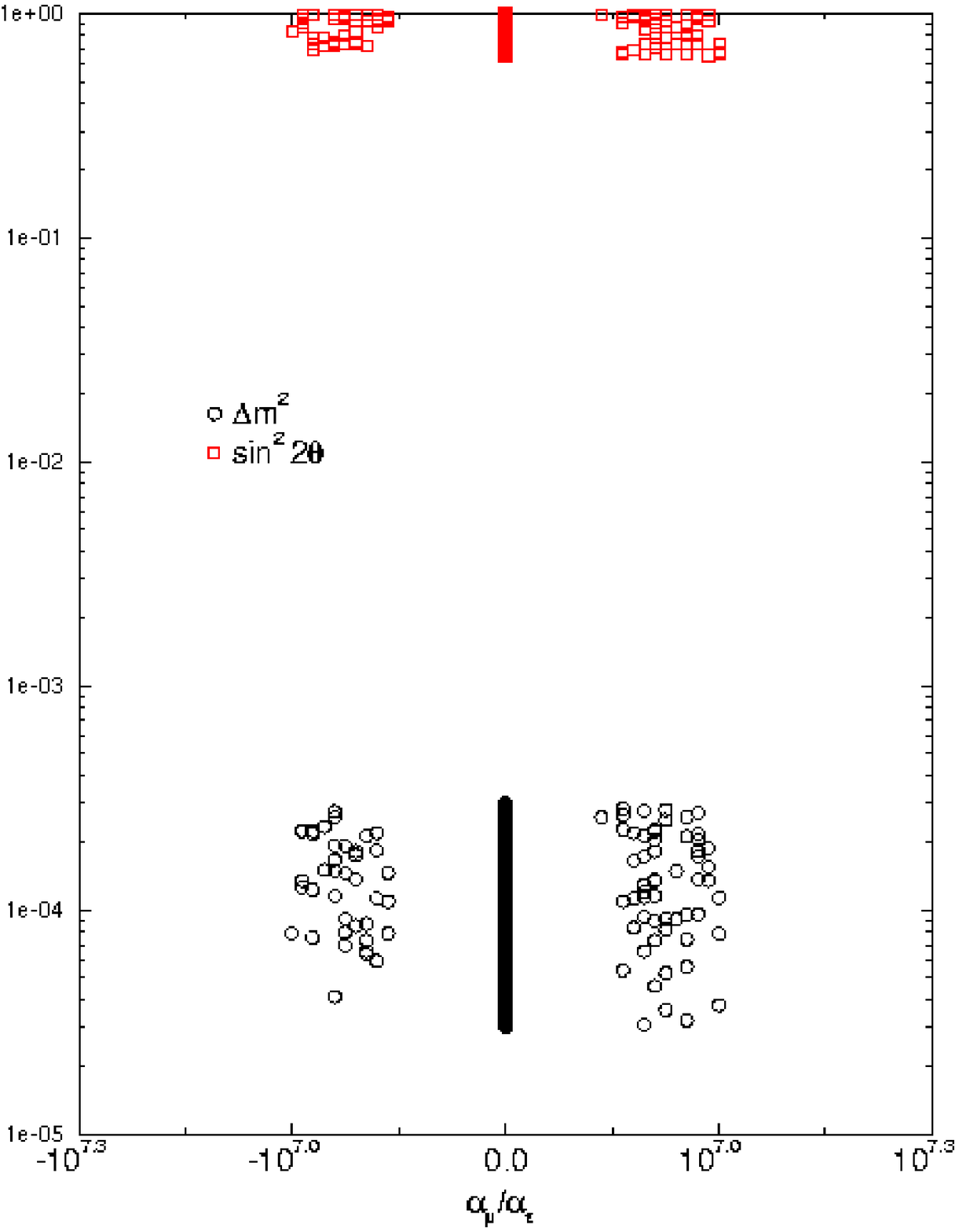}}
\centerline{\hspace{1.5cm} (a) \hspace{7.5cm} (b)}
\vspace{-0.1cm}
\centerline{\hspace{-3.3mm}
\epsfxsize=8cm\epsfbox{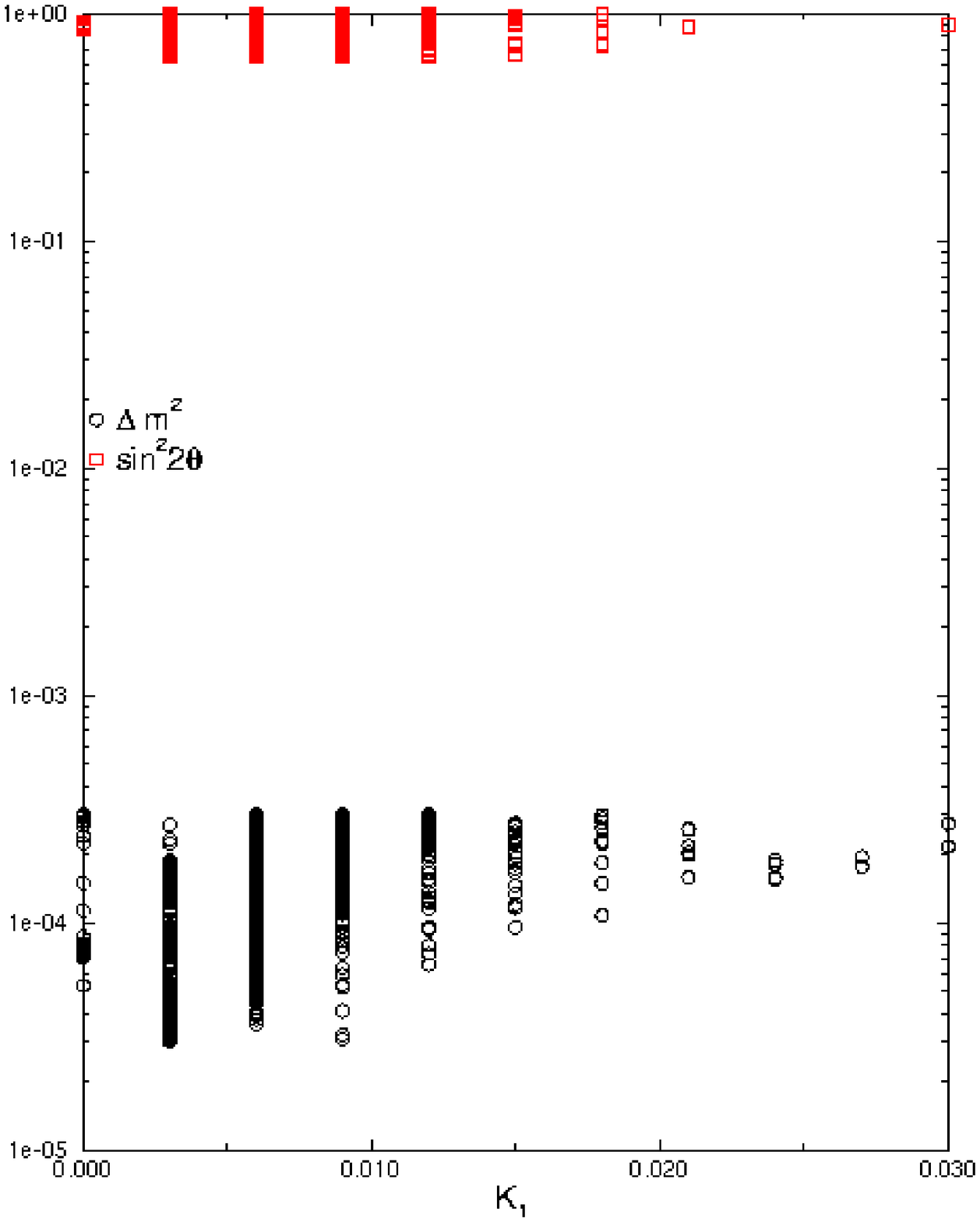}
\hspace{-0.1cm}
\epsfxsize=8cm\epsfbox{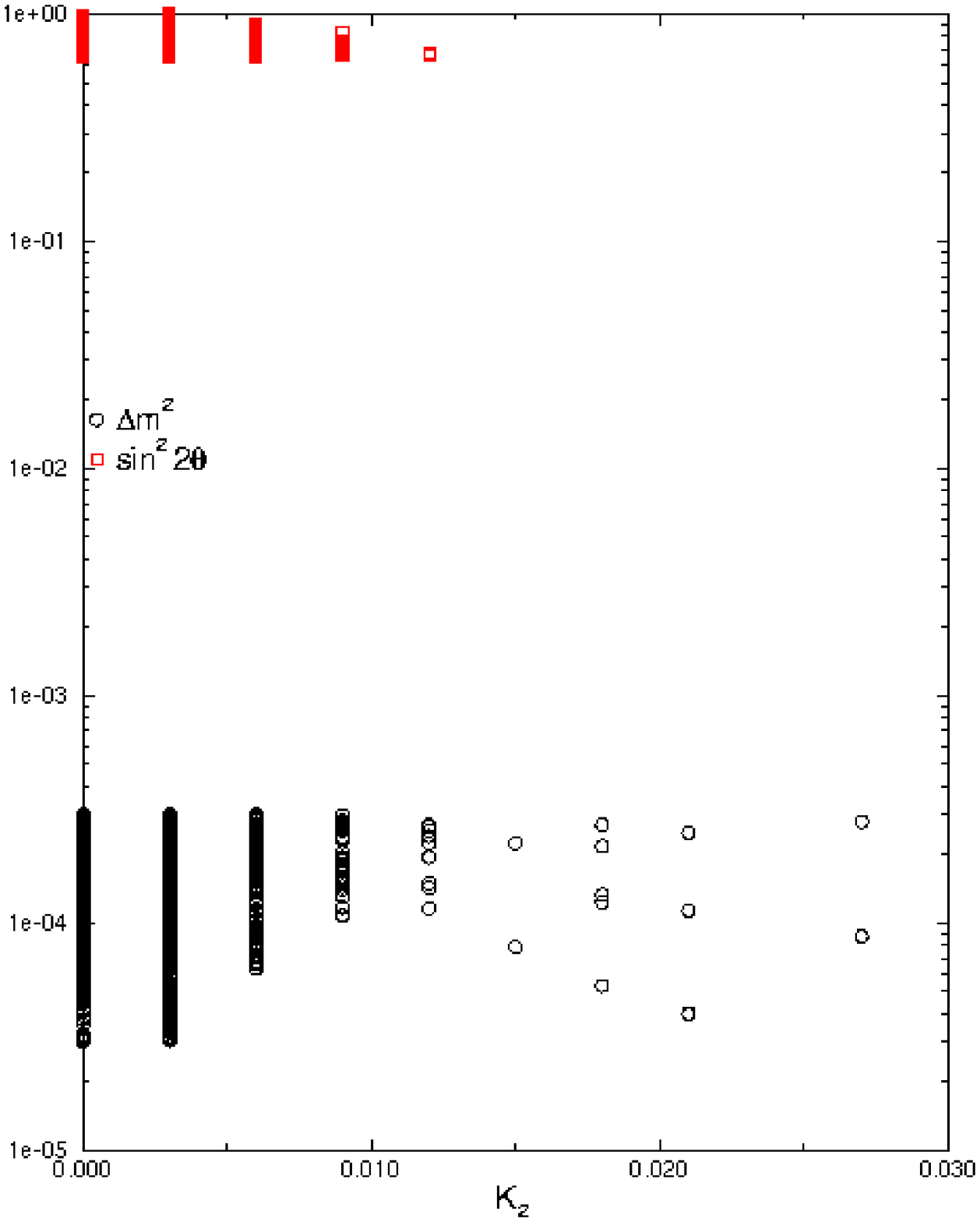}}
\centerline{\hspace{1.5cm} (c) \hspace{7.5cm} (d)}

\vspace*{-0.1cm}
\caption{The solar parameters 
$\Delta m^2_{12}$ and $\sin^2 2\theta_{12}$
 as functions of the different R-parity parameters $\alpha_e$ (a), 
  $\alpha_\mu/\alpha_\tau$ (b), $K_1$ ( $\propto\lambda^2$) (c) and 
   $K_2$ ($\propto \lambda'^2$) (d), in the case of a combined fit with MSW-LMA, SuperK and CHOOZ
   constraints. }\protect\label{LMA-x-sun}

\end{figure}

\vspace*{-0.3cm}

\begin{figure}[hbt]


\centerline{\hspace{-3.3mm}
\epsfxsize=8cm\epsfbox{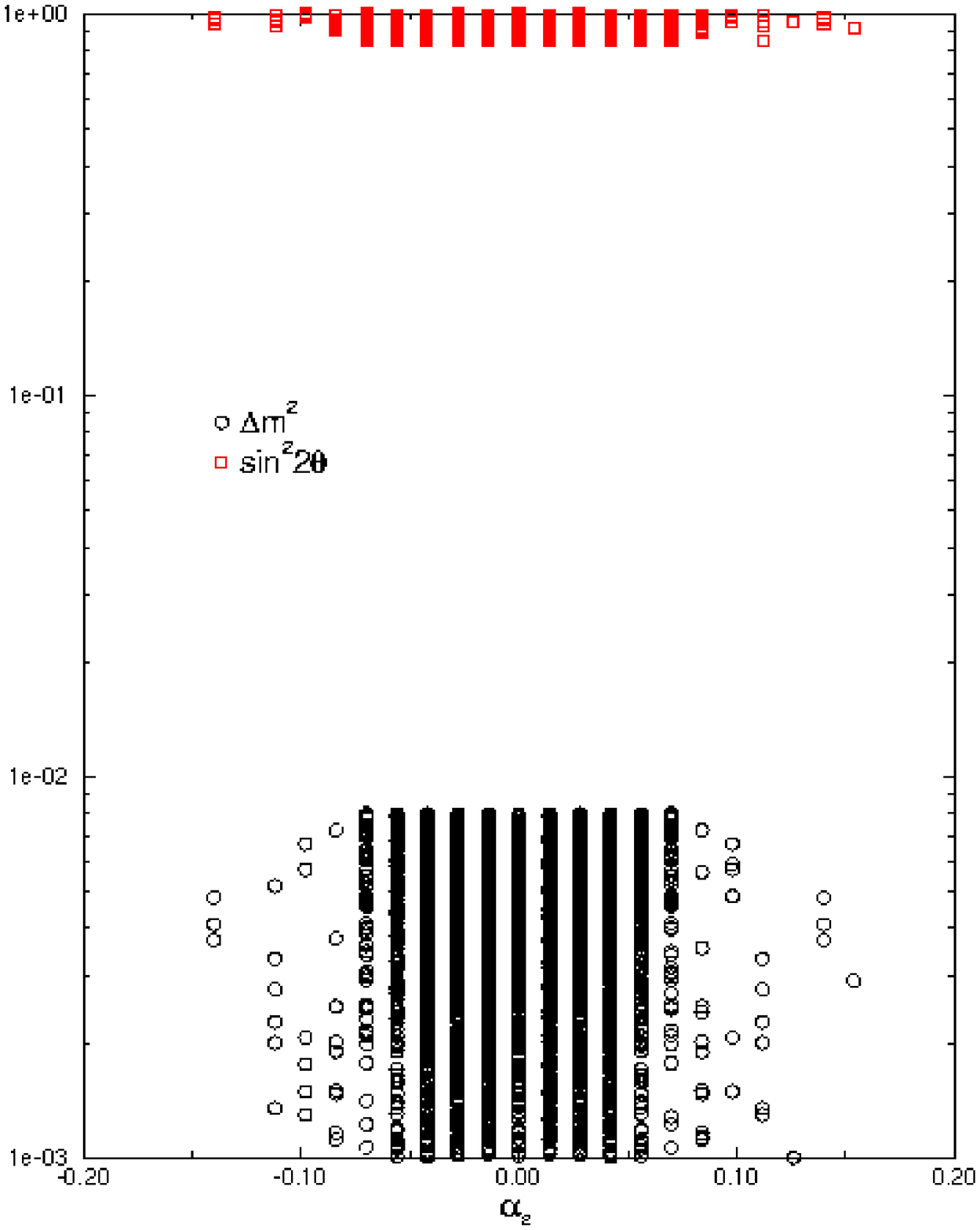}
\hspace{-0.1cm}
\epsfxsize=8cm\epsfbox{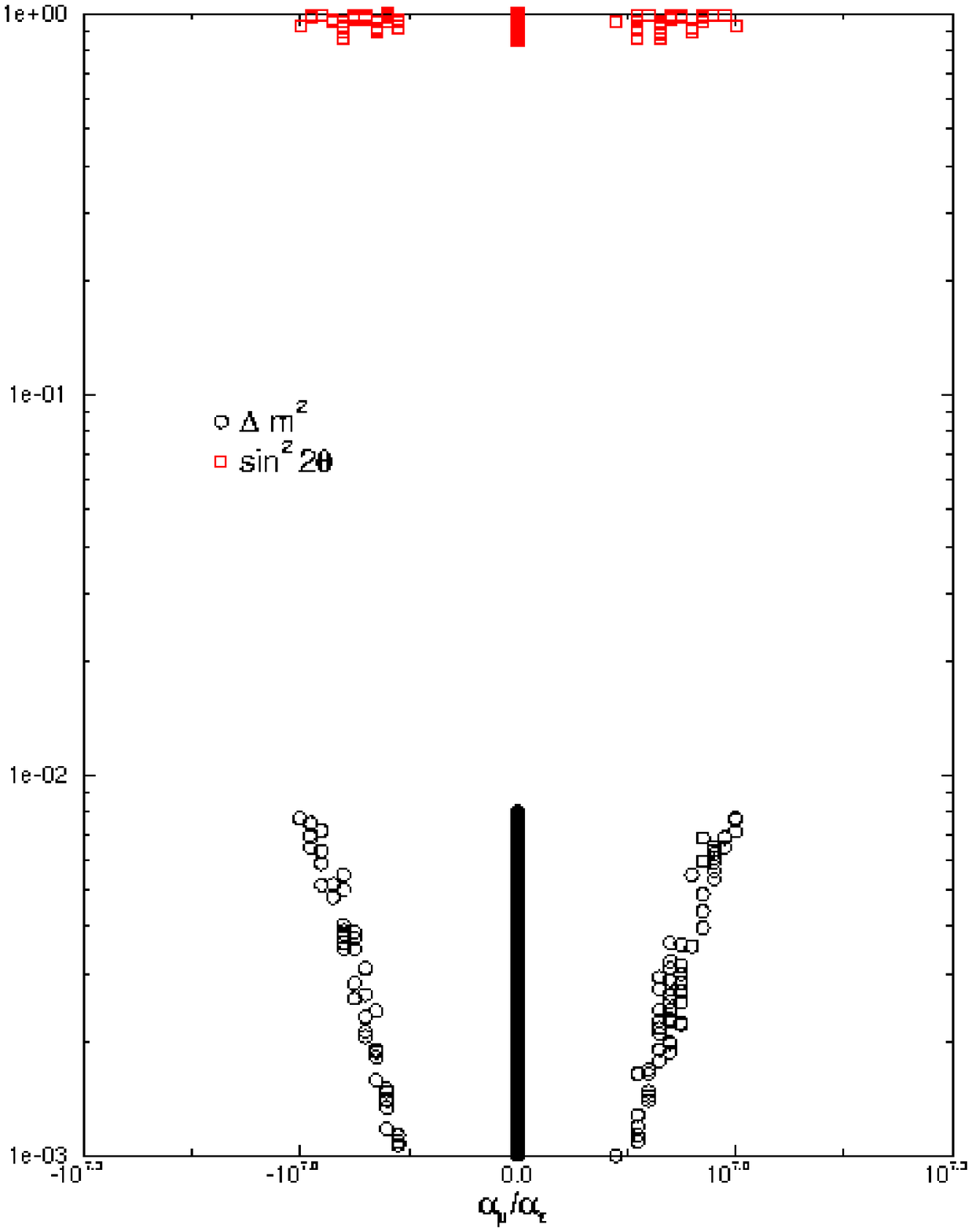}}
\vspace*{-0.3cm}
\centerline{\hspace{1.5cm} (a) \hspace{7.5cm} (b)}
\vspace{.0001cm}
\vspace{-0.1cm}
\centerline{\hspace{-3.3mm}
\epsfxsize=8cm\epsfbox{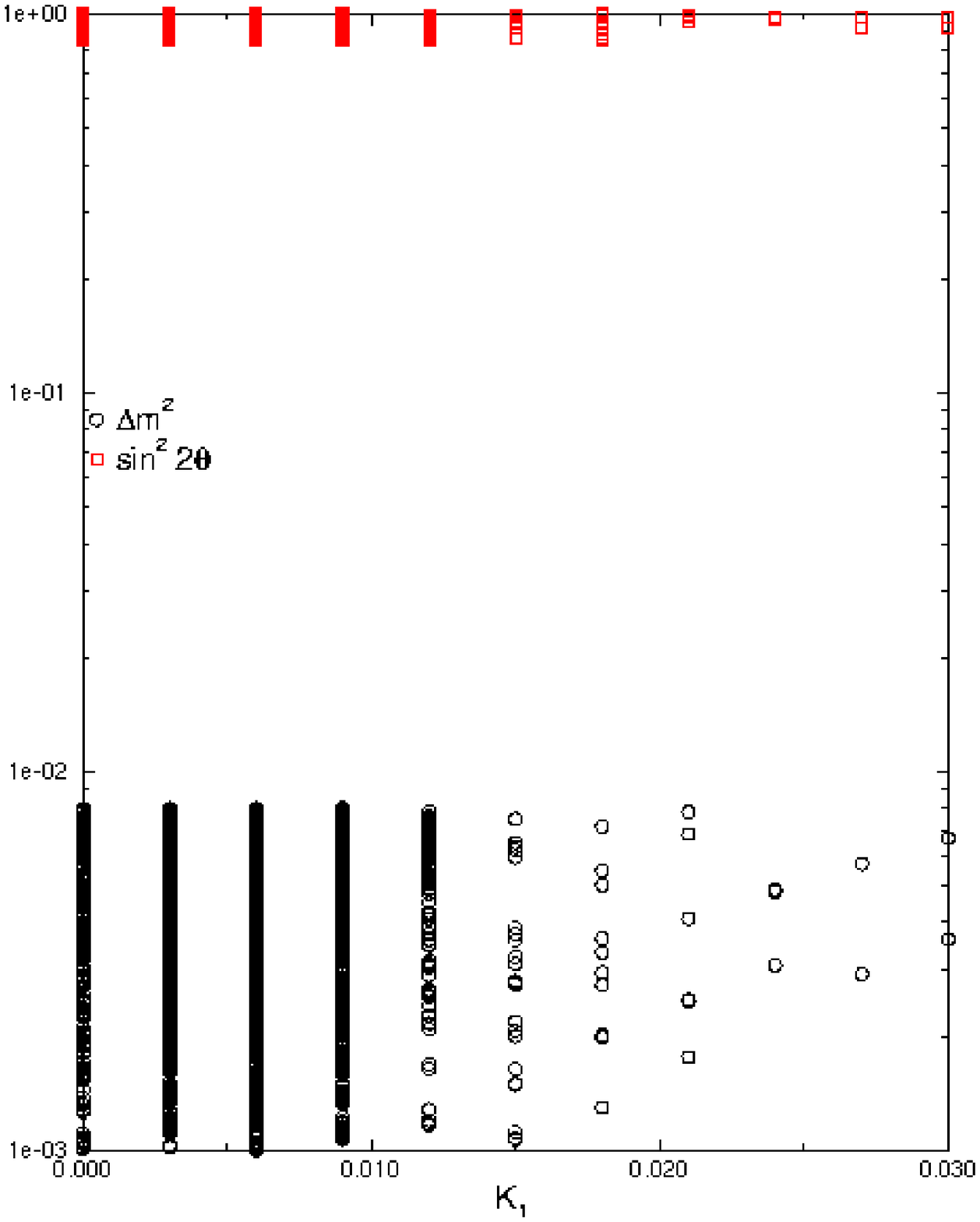}
\hspace{-0.1cm}
\epsfxsize=8cm\epsfbox{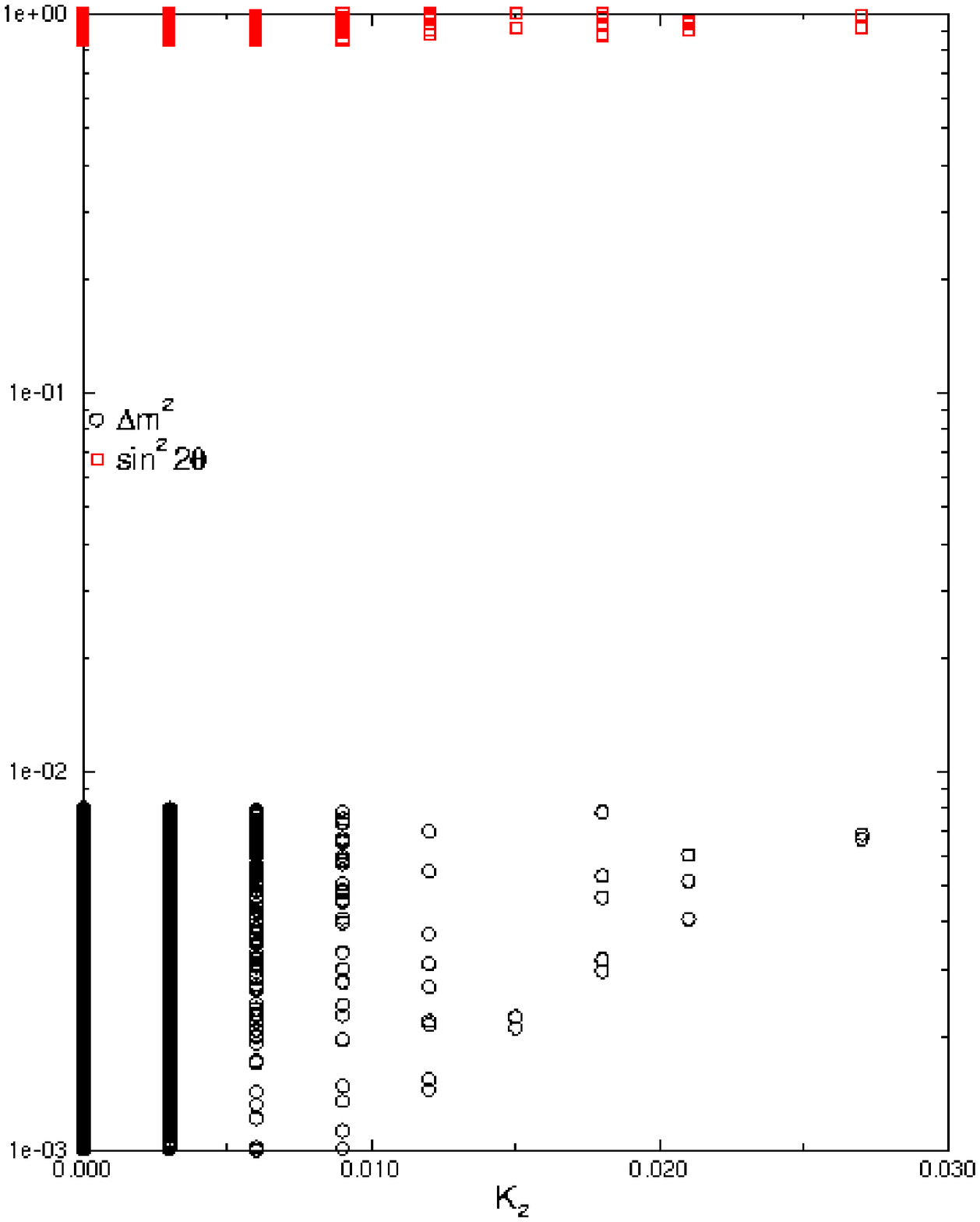}}
\centerline{\hspace{1.5cm} (c) \hspace{7.5cm} (d)}

\vspace*{-0.1cm}
\caption{The atmospheric parameters $\Delta m^2_{13}$ and 
$\sin^2 2\theta_{23}$
 as functions of the different ${\mathrm {R}}_{\mathrm{P}}$ parameters $\alpha_e$ (a), 
  $\alpha_\mu/\alpha_\tau$ (b), $K_1$ ($\propto \lambda^2$) (c) and 
   $K_2$ ($\propto\lambda'^2$) (d), in the case of a combined fit with MSW-LMA, SuperK and CHOOZ
   constraints.}\protect\label{LMA-x-atm}

\end{figure}

\vspace*{-0.3cm}

\begin{figure}[hbt]


\centerline{\hspace{-3.3mm}
\epsfxsize=8cm\epsfbox{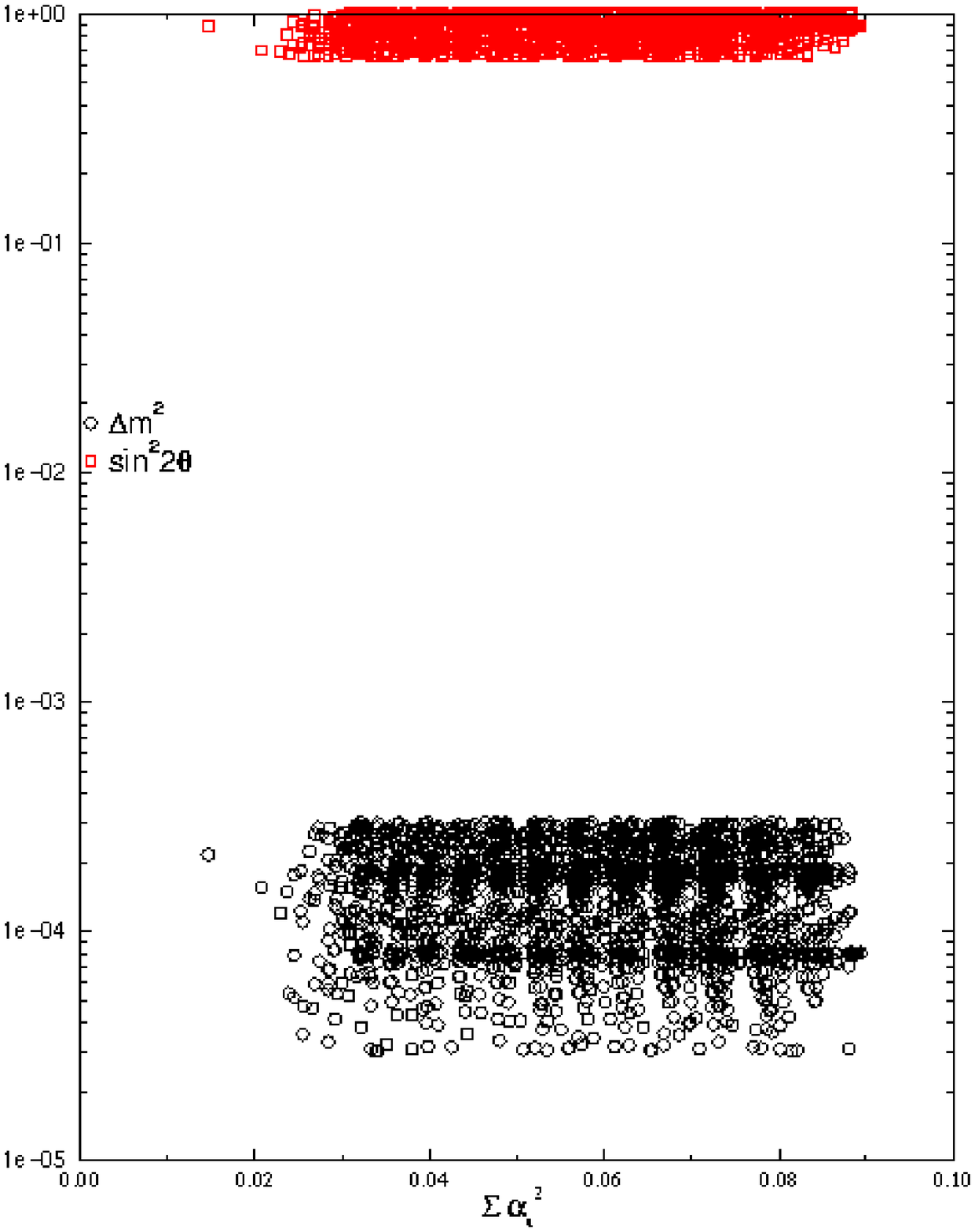}
\hspace{-0.1cm}
\epsfxsize=8cm\epsfbox{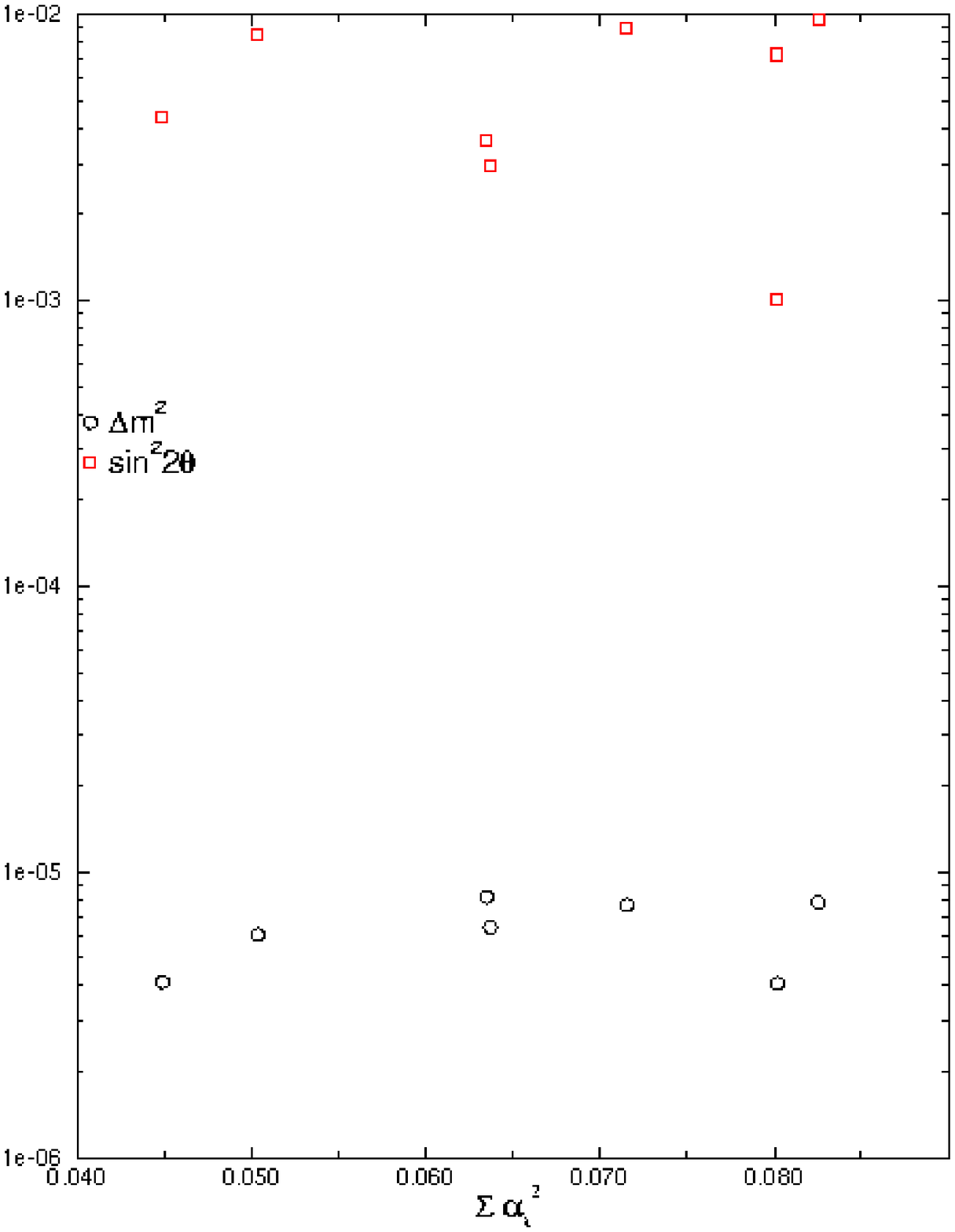}}
\vspace*{-0.3cm}
\centerline{\hspace{1.5cm} (a) \hspace{7.5cm} (b)}
\vspace{.0001cm}
\centerline{\hspace{-3.3mm}
\epsfxsize=8cm\epsfbox{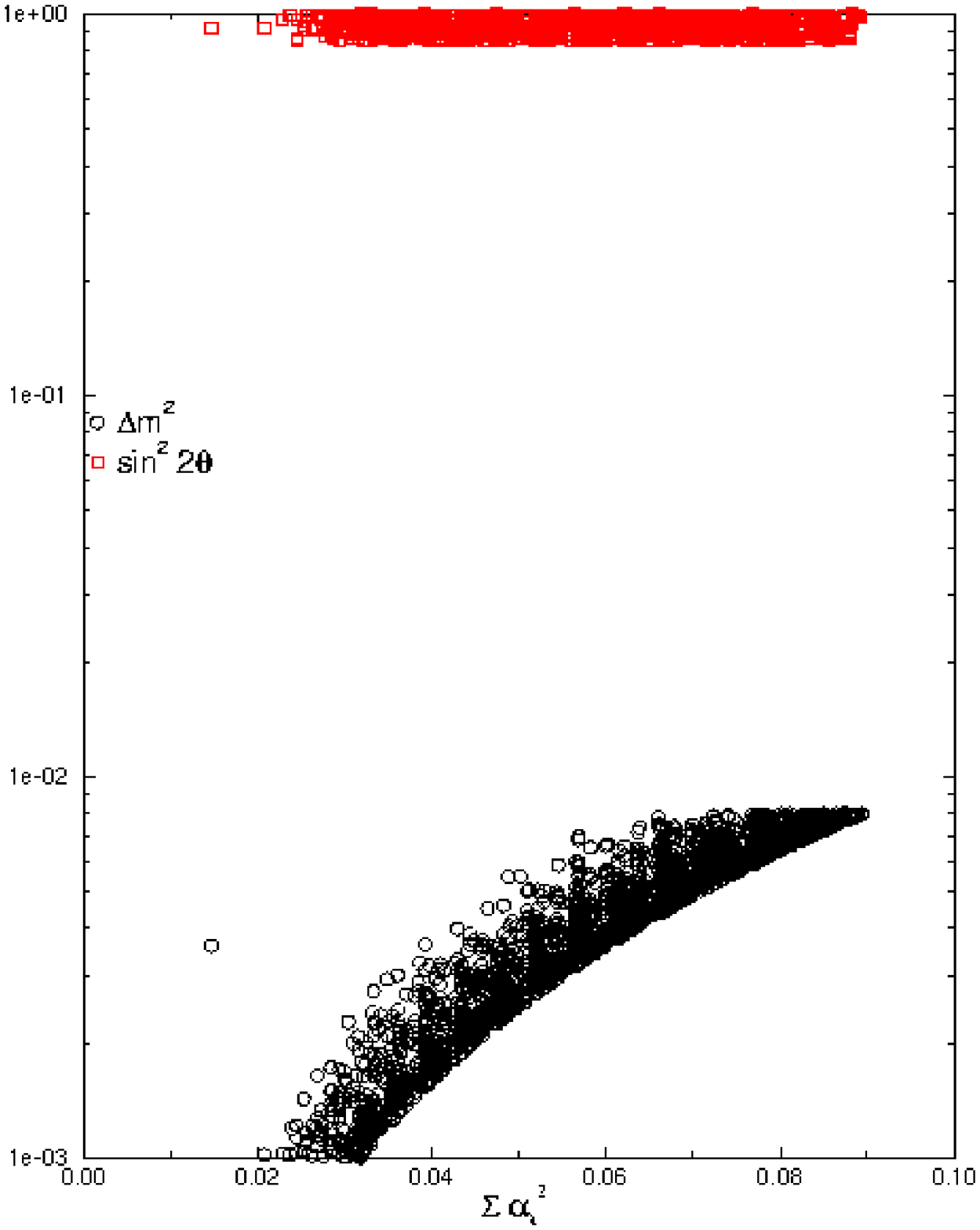}
\hspace{-0.1cm}
\epsfxsize=8cm\epsfbox{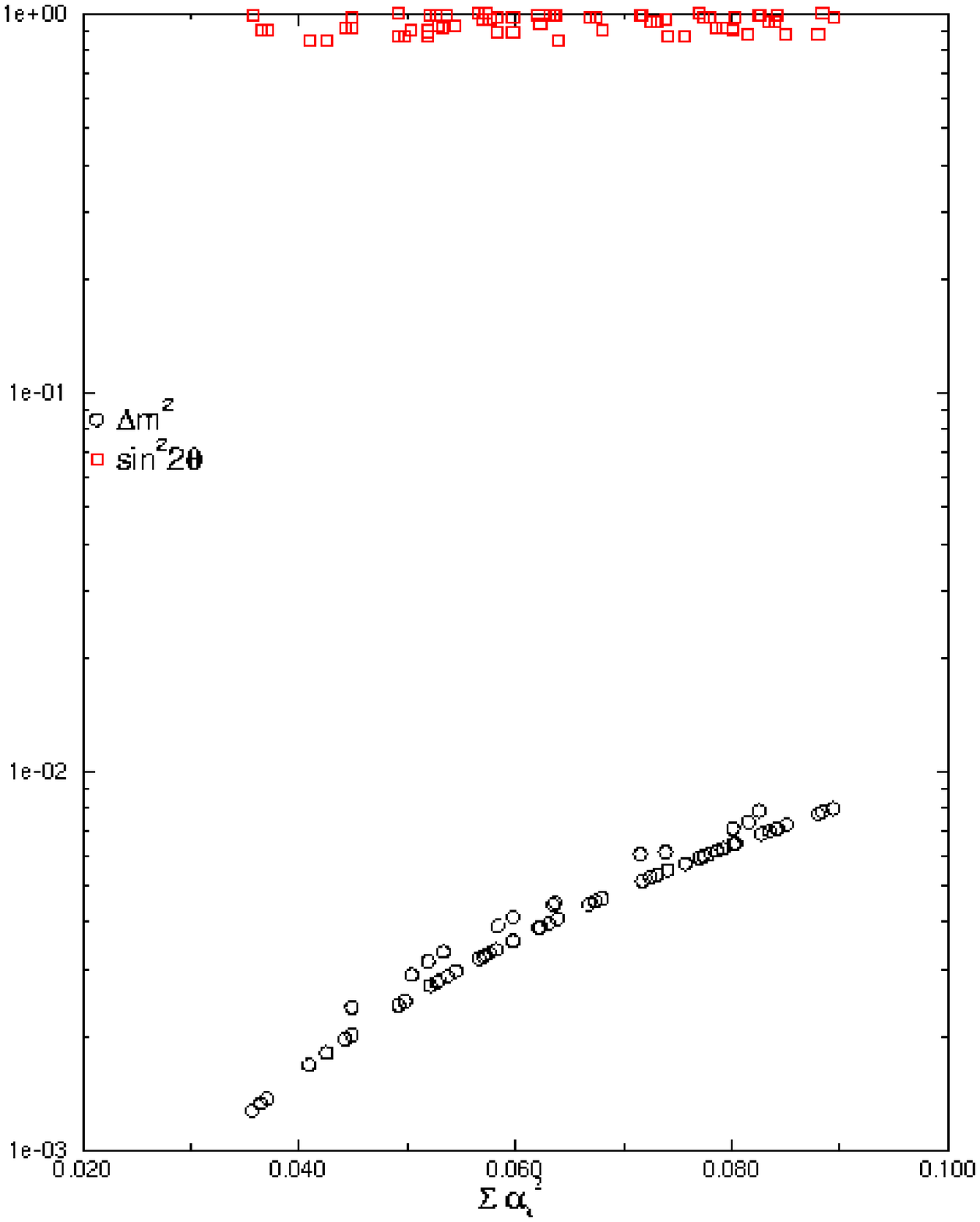}}
\centerline{\hspace{1.5cm} (c) \hspace{7.5cm} (d)}
\vspace*{-0.1cm}

\caption{The atmospheric 
($\Delta m^2_{13}$ and $\sin^2 2\theta_{23}$) (in (a))
and solar  parameters ($\Delta m^2_{12}$ and $\sin^2 2\theta_{12}$) (in (c))
 as functions of $\alpha^2= \alpha_e^2 +\alpha_\mu^2+\alpha_\tau^2$ in the case
 of MSW-LMA solution  (a) and (c) and in the case of MSW-SMA (b) and (d). SuperK and CHOOZ 
  constraints are also imposed.}
 \protect\label{LMA-sum}

\end{figure}

 \section{Discussions and conclusions}

Our results from the general analysis show that there are many  regions
in  parameter space that can accommodate simultaneously the MSW-LMA
and atmospheric oscillation solutions together with the CHOOZ
constraint, see figs.  2 and 3. In comparison, we see  from fig. \ref{SMACSK}a that not
all of the allowed region in the ($\Delta m^2$, $\sin^{2} 2\theta$) plane
is covered when  MSW-SMA and the atmospheric 
oscillations  solutions together with the CHOOZ constraints are imposed.
 Moreover,  the region of parameter space for which the atmospheric, CHOOZ and vacuum oscillation
 constraints are satisfied has very few solutions.
In contrast, we find that for the specific case of the MSSM with R-parity violation
there are no regions in parameter which can accommodate simultaneously the vacuum
solar solution and all other constraints.

As far as the possible spectra are concerned, we  clearly see that  for the
 combined
MSW-SMA, CHOOZ and SuperK constraints both the hierarchical 
and pseudo-Dirac spectra are more commonly 
present. 
There are also a few values of  parameter space that  present
a degenerate spectrum.
For a combination of MSW-LMA, CHOOZ and SuperK constraints, 
the three possible spectra are very common.
As expected, for the combined vacuum oscillation solution,  CHOOZ and SuperK 
constraints
 only pseudo-Dirac spectra are present.
In reference to figs. 4, 5b and 6b our  conclusions hold and the bounds 
mentioned in the previous section are still maintained
when we study the spectrum in terms of the other parameters $M_{ij}\neq M_{11}$.

It is clear that increasing the range of variation of our parameters changes
the values of the eigenvalues of the mass matrix. In fact, larger values of
the inputs make it impossible to satisfy the vacuum oscillation constraints; 
eventually the same will occur with the MSW-SMA and MSW-LMA solutions.
It is really the combination of constraints that solve both atmospheric and solar
anomalies which limit the allowed ranges. 
An important conclusion is that, after obtaining a simultaneous solution
for the solar and atmospheric anomalies and the CHOOZ constraint,  
  the effective mass constraint and the cosmological constraint
are always satisfied. We have also indicated previously when a stronger bound
can be placed on these quantities. 

Some of the previous work regarding neutrino masses in this specific 
 model we consider have tried to fit  only one of the $\Delta m^2$ to either
 the atmospheric or one of the solar solutions. From our analysis
we have seen that fitting only one of the neutrino anomalies
is straightforward. What is much harder to obtain is a region
in parameter space that satisfy the combined fits.
Other authors have also studied the possibility of fitting
both atmospheric and solar anomalies, although their approaches
are somewhat different from ours. They consider either
 supergravity-inspired models or exclusively bilinear contributions, or exclusively trilinear
contributions.

 The usual bounds \cite{dreiner,bhatta2} cited for the trilinear couplings 
 $\lambda_{133}$ and $\lambda'_{133}$ are placed when only one of these two 
 R-parity-violating couplings are considered to be 
present in the theory. The bound is implemented by using the upper bound on the electron
neutrino mass from the
$^3 {\mathrm{H}}\ \beta$-decay spectrum. Having only one of
 the trilinears  corresponds of course to having a single
non-zero matrix element in eq. (\ref{MRp}), given by $K_{1}$ or $K_{2}$ depending on whether the trilinear coupling is  $\lambda_{133}$ or $\lambda'_{133}$,
respectively. We see  that in the special limits 
we have considered on R-parity-conserving 
SUSY parameters our bounds from eq. (\ref{num1l}) agree with the previous bounds.  
The overall conclusion is that the bounds are very strong on the R-parity  violating
 parameters.
We have shown that the simultaneous inclusion of many of these couplings 
can constrain them much more than other bounds arising from other low-energy or high-energy
processes \cite{dreiner,bhatta2}. 
The overall conclusion is that the bounds are very strong on the R-parity
violating parameters.

Figures 10-12 illustrate the different dependences of the solar and atmospheric
parameters on the R-parity couplings. It is clear that for large values of the parameters
there are no solutions. The discreteness of the allowed values of the parameters  is
 due in part to the step we use in our scan of the parameter space. We also see, of course,
  that
the dependence on a single parameter is complicated, since  many solutions can 
be found for the same value of the variable. This is due to the large number of
different parameters involved.

The general global interpretation of the experimental data is that the SuperK
seriously challenges MSW-SMA  and Vacuum Oscillations solutions  \cite{Chang-kee}. This is in
agreement with the analysis we have done. Indeed, the majority of the solar anomaly solutions
we have that also satisfy both SuperK and CHOOZ  are MSW-LMA.   
The future experiments SNO, BOREXINO, ICARUS, HERON with solar neutrinos as well as 
 KAMLAND, K2K, MINOS, BooNE..., and many others, will provide additional decisive 
 information to pin down the neutrino mysteries.

\vspace{0.5cm}
\noindent

{\bf Acknowledgments } We thank S. Davidson for very interesting and
useful discussions.

\clearpage
\appendix 
\centerline{\bf Charged and Neutral Fermion Mass Matrices}

\section{Charged colour-singlet mass matrix}\label{appendixa1}
Written in the basis
$\Psi= (-i \lambda_{-}, \psi_{L_{o}}^{2},\psi_{L_{1}}^{2},\psi_{L_{2}}^{2},
\psi_{L_{3}}^{2})$, the matrix  

 \bea{\cal M}_{c}^{tree}=
 \left(
 \begin{array}{ccccc}
M_{2} & {g_{2} v \sin\beta\over \sqrt2} &0&0&0\\
{g_{2} v \cos\beta\over \sqrt2}& \mu_{o} &0&0&0\\
0&\mu_{1}&m_{1}&0&0\\
0&\mu_{2}&0&m_{2}&0\\
0&\mu_{3}&0&0&m_{3}
  \end{array}
 \right)\ ,
\label{mcharginos}
  \eea
can be diagonalized explicitly to obtain the charginos and charged-lepton tree-level masses.
However, our analysis of the neutrino spectrum requires the $\mu_{i}$'s to be very small, thus their
effect on the eigenvalues  and corresponding mass eigenstates will be tiny.
It is a good approximation for small R-parity violation to neglect the  RPV contribution to the masses
in the charged leptons  and charginos \cite{kongetal}.

 \section{Tree-level neutral colour-singlet mass matrix}\label{appendixa2}

 \bea{\cal M}_{N}^{\mathrm{tree}} = \left( \begin{array}{cc}
M& \xi^{T}\\
\xi& m_{\nu}^{o}
\end{array}
\right)=
 \left(
 \begin{array}{ccccccc}
M_{1}&0 & {g_{1} v \sin\beta\over \sqrt2} & -{g_{1} v \cos\beta\over \sqrt2}& 0&0&0\\
0&M_{2}&{g_{2} v \sin\beta\over \sqrt2}&-{g_{2} v \cos\beta\over \sqrt2} &0&0&0\\
 {g_{1} v \sin\beta\over \sqrt2}& -{g_{2} v \sin\beta\over \sqrt2}&0&\mu_{o}&-\mu_{1}&-\mu_{2}&-\mu_{3}\\
- {g_{1} v \cos \beta\over \sqrt2}&{g_{1} v \sin\beta\over \sqrt2}&-\mu_{o}&0&0&0&0\\
0&0&-\mu_{1}&0&0&0&0\\
0&0&-\mu_{2}&0&0&0&0\\
0&0&-\mu_{3}&0&0&0&0\\
  \end{array}
 \right)\ ,
\label{mneutralinos}
  \eea

In the limit where the $\mu_{i}$'s are small, the effective neutrino mass matrix
has the ``see-saw'' structure 

\begin{equation}
{\cal M}_{\nu}^{\mathrm{tree}} = - \xi M^{-1} \xi^{T} + m_{\nu}^{o}\ .
\label{effecMass}
\end{equation}
The assumption we make so as to include the  one-loop corrections is that only 
the corrections to $m_{\nu}^{o}$ need  be considered, thus giving the
one-loop-corrected expression for the $3\times 3$ neutrino mass matrix
\cite{nowakowski}, \cite{Bednyakov}, \cite{Kong}:

\begin{equation}
{\cal M}_{\nu} = - \xi M^{-1} \xi^{T} + {\cal M}_{\nu}^{\mathrm{loop}}\ ,
\label{effecMass1}
\end{equation}
where $ {\cal M}_{\nu}^{\mathrm{loop}} =  m_{\nu}^{o} + m_{\nu}^{\mathrm{loop}}$. 
The expression for $ m_{\nu}^{\mathrm{loop}}$ is given
in the next subsection. The elements of 
the matrix $M^{\mathrm{tree}}_{\nu}= \xi M^{-1} \xi^{T}$, 
which is parametrized in terms of the $\alpha_{i}$'s of eq. (\ref{Malphas}) are  given
by 
\begin{equation}
M^{\mathrm{tree}}_{\nu_{ij}} = g_{2}^{2}{(M_{1} + \tan^2\theta_{W} M_{2})\over 4 \det M}\mu_{i}\mu_{j} v_{1}^{2},
\label{treeelem}
\end{equation}
as we are working in the basis where the slepton vacuum 
expectation value is zero, $v_{1} = v_{d} = v \cos\beta$. 
The basis-invariant expression corresponds to substituting $\mu_{i}\mu_{j} v_{1}^{2} $ with
$(\mu <v_{i}> - \mu_{i} v_{1})(\mu \langle v_{j}\rangle - \mu_{j} v_{1})$, where 
$\langle v_{i}\rangle$ is the vacuum expectation value of the slepton fields
$L_{i}$\footnote{For a general discussion of basis-independent parametrizations of
R-parity violation, see refs. \cite{davidson}.}.

 \section{Loop-mass matrix}\label{appendixa3}
 Expliciting the indices in eq. (\ref{loopm}) leads to the following symmetric 
 matrix elements
 of which  we keep only the relevant contributions due to the mass hierarchy in the
  charged-lepton sector 
 $m_e\ll m_\mu\ll m_\tau$ and in the down-quark sector $m_d\ll m_s\ll m_b$:
 
  \bea
 &&m_{11}^{\mathrm{loop}}={X\over 16 \pi^2} \left({f(x_\ell) \over M^{2}_2} \left(
 \lambda^2_{133}m_{\tau}^2  \right)+ 3{f(x_q) \over M^{q2}_2}\left(
 \lambda'^2_{133}m_{b}^2 +  \lambda'_{123}\lambda'_{132} m_s m_b    \right) \right)
\label{loopm11}\\
 &&m_{12}^{\mathrm{loop}}={X\over 16 \pi^2} \left({f(x_\ell) \over M^{2}_2} \left(
 \lambda_{133}\lambda_{233}m_{\tau}^2  \right)+ 3{f(x_q) \over M^{q2}_2}\left(
 \lambda'_{133}\lambda'_{233}m_{b}^2 + ( \lambda'_{132}\lambda'_{223}+
 \lambda'_{123}\lambda'_{232})m_s m_b    \right) \right)
\label{loopm12}\\
&&m_{13}^{\mathrm{loop}}={3\ X\over 16 \pi^2}  {f(x_q) \over M^{q2}_2}\ 
 \lambda'_{133}\lambda'_{333}m_{b}^2      
\label{loopm13}\\
&&m_{22}^{\mathrm{loop}}={X\over 16 \pi^2} \left({f(x_\ell) \over M^{2}_2} \left(
 \lambda^2_{233}m_{\tau}^2  \right)+ 3{f(x_q) \over M^{q2}_2}\left(
 \lambda'^2_{233}m_{b}^2 +  \lambda'_{223}\lambda'_{232} m_s m_b    \right) \right)
\label{loopm22}\\
&&m_{23}^{\mathrm{loop}}={3\ X\over 16 \pi^2}  {f(x_q) \over M^{q2}_2}\ 
 \lambda'_{233}\lambda'_{333}m_{b}^2      
\label{loopm23}\\
&&m_{33}^{\mathrm{loop}}={3\ X\over 16 \pi^2}  {f(x_q) \over M^{q2}_2}\ 
 \lambda'^2_{333} m_{b}^2      
\label{loopm33}
\eea


\begin{thebibliography}{99}


 \bibitem{ska} Y. Fukuda et al., Super-Kamiokande Collaboration, Phys. Lett. {\bf B 433}
 (1998) 9;
 Phys. Lett. {\bf B 436} (1998) 33; Phys.Rev.Lett. {\bf 81} (1998) 1562.
 \bibitem{imb} R. Becker-Szendy et al., IMB Collaboration, Nucl. Phys. {\bf B} 
 (Proc. Suppl.) {\bf 38} (1995) 331.
 \bibitem{soudan} W.W.M. Allison et al. Soudan-2 Collaboration,
  Phys. Lett. {\bf B391} (1997) 491, Phys. Lett. {\bf B449} (1999) 137. 
 \bibitem{ka} Y. Fukuda et al., Kamiokande Collaboration, Phys. Lett. 
 {\bf B 335}
 (1994) 237.
 \bibitem{davis}R. Davis et al., Phys.Rev.Lett. {\bf 21} (1968) 1205; 
 B.T. Cleveland et al., Astrophys. J. {\bf 496} (1998) 505.
 
 \bibitem{gallex} W. Hampel et al., Phys. Lett. {\bf B 388} (1996) 384.
 \bibitem{sage}D.N. Abdurashitov et al., Phys. Rev. Lett. {\bf B 77} (1996) 
 4708; astro-ph/9907131.
 \bibitem{ks}K. S. Hirata et al., Kamiokande Collaboration, Phys. Rev. Lett. {\bf 77}
 (1996) 1683. 
  \bibitem{sks}Y. Fukuda et al., Super-Kamiokande Collaboration, Phys.Rev.Lett. 
  {\bf 81} (1998) 1158.
 \bibitem{macro}M. Ambrosio et al., MACRO  Collaboration, Phys. Lett. {\bf B 434}
 (1998) 451.
 \bibitem{lsnd}C. Anthanassopoulos et al., LSND Collaboration, Phys. Rev. Lett. 
 {\bf B 81}
 (1998) 1774.
 


\bibitem{coleman-glashow} S. Coleman and S.L. Glashow, Phys. Rev. {\bf D59} 116008  and references therin.
\bibitem{halprin} A. Halprin and H.B. Kim. KIAS-P99028, 
preprint hep-ph/9905301 and references therin.

\bibitem{pakvasa}S. Pakvasa, hep-ph/9905426.
\bibitem{fogli2}G.L. Fogli, E. Lisi, A. Marrone and G. Scioscia, 
 Phys.Rev. {\bf D60} 053006 and  Phys.Rev. {\bf D59} 117303. 
\bibitem{karmen} B. Zeitnitz et al.,KARMEN Collaboration 
  Prog.Part.Nucl.Phys. {\bf 40} (1998) 169.

\bibitem{neutrinos}see for example and refs. therein:
 R.Barbieri, J. Ellis and M.K. Gaillard,  Phys. Lett. {\bf B90} (1980) 249;
M. Gell-Mann, P. Ramond and R. Slansky, 
{\em Proceedings of the Supergravity Stony Brook Workshop}, New York 1979, 
eds. P. Van Nieuwenhuizen and D. Freedman; T. Yanagida, 
{\em Proceedinds of the Workshop on Unified Theories and Baryon Number in the Universe}, 
Tsukuba, Japan 1979, ed.s A. Sawada and A. Sugamoto; 
S. Dimopoulos, L.J. Hall and S. Raby, Phys. Rev. Lett. {\bf 68} (1992) 1984; 
G.K. Leontaris et al,  Phys. Rev. {\bf D53}  (1996) 6381;
 S. Lola and J. D. Vergados,  Prog. Part. Nucl. Phys.{\bf 40} (1998) 71; 
 G. Altarelli and F. Feruglio, hep-ph/9905536 and hep-ph/9907532;
J.A. Casas et al, hep-ph/9905381 and hep-ph/9906281;
 M. Carena et al, hep-ph/9906362; 
 P. Langacker,  Nucl. Phys. Proc. Suppl. {\bf 77} (1999) 241; 
 Z. Berezhiani and A. Rossi, JHEP {\bf 9903} (1999) 002.
\bibitem{barbieri} R. Barbieri, L.J. Hall, D. Smith, A. Strumia and N. Weiner, 
JHEP {\bf 9812} (1998) 017.
\bibitem{roulet}G. Gelmini and E. Roulet,
  Rept. Prog. Phys. {\bf 58} (1995) 1207.
\bibitem{kayser} B. Kayser, F. Gibrat-Bebu and F. Perrier, {\em The Physics of
Massive Neutrinos}, Wolrd Scientific, Singapore 1989.
\bibitem{bilenky}S.M. Bilenkii, C. Giunti, W. Grimus, hep-ph/9812360. 


\bibitem{valle}J.W.F. Valle, hep-ph/9809234.
\bibitem{chooz} M. Apollonio et al., CHOOZ Collaboration, Phys. lett. {\bf B 420}
 (1998) 397.
\bibitem{G-G} H. Georgi and S.L. Glashow, hep-ph/9808293.
\bibitem{M-H} M. G\"unther et al., Phys. Rev.  {\bf D55} (1997) 54, 
   Phys. lett. {\bf B 407} (1997) 219.
\bibitem{M-H-2} L. Baudis et al., Phys. Lett. {\bf B 407}
 (1997) 219.
\bibitem{cosmo} R.A.C. Croft, W. Hu and R. Dav\'e, astro-ph/9903335, and see
   references therein.
 \bibitem{troitsk} V. M. Lobashev,  
  to be published in the proceedings of 17th 
International Workshop on Weak Interactions and Neutrinos (WIN 99), 
Cape Town, South Africa, 24-30,Jan 1999; Prog.Part.Nucl.Phys.{\bf 40} (1998) 337.
\bibitem{mainz} H. Barth, Prog.Part.Nucl.Phys.{\bf 40} (1998) 353, 
 J. Bonn, to be published in the proceedings of 17th 
International Workshop on Weak Interactions and Neutrinos (WIN 99), 
Cape Town, South Africa, 24-30,Jan 1999. 
 \bibitem{cosmological} G. Gelmini, 
to be published in the proceedings of 17th 
International Workshop on Weak Interactions and Neutrinos (WIN 99), 
Cape Town, South Africa, 24-30, Jan 1999, hep-ph/9904369.
 \bibitem{Fogli} G.L. Fogli, E. Lisi, A. Marrone and G. Scioscia, 
 Phys. Rev. {\bf D59} 33001.
 \bibitem{Barger}V. Barger, T.J. Weiler and K. Whisnat, 
 Phys. Lett. {\bf B440}  (1998)1.
\bibitem{concha} M. C. Gonzalez-Garcia, H. Nunokawa, O.L.
 G. Peres and J.W.F. Valle,  Nucl. Phys. {\bf B543}  (1999) 3.
 \bibitem{bahcall} J.N. Bahcall, P.I. Krastev and A.Yu. Smirnov, 
Phys. Rev. {\bf D58} 96016.
 \bibitem{hata} N. Hata and P. Langacker, Phys. Rev. {\bf D56} (1997) 6117.
 \bibitem{petcovs} S.T. Petcov, to be published in the proceedings of 17th 
International Workshop on Weak Interactions and Neutrinos (WIN 99), 
Cape Town, South Africa, 24-30, Jan 1999,  hep-ph/9907216.
 \bibitem{C-K} L.L. Chau and W.Y. Keung, Phys.Rev.Lett. {\bf 53} (1984) 1802.
\bibitem{Godbole} R. Godbole, P. Roy and X. Tata,  Nucl. Phys. {\bf B401}  (1993) 67.
\bibitem{haber} Y. Grossman,  H.E. Haber  Phys.Rev. {\bf D59} 093008. 
\bibitem{nardi} E.Nardi, Phys. Rev. {\bf D55} (1997) 5772.
\bibitem{banks} T. Banks, Y. Grossman, E. Nardi and Y. Nir,  Phys. Rev. {\bf D52} (1995) 5319.

\bibitem{nowakowski} M. Nowakowski and A. Pilaftsis, 
Nucl. Phys. {\bf B461}  (1996) 19.

\bibitem{Borzumati} F.M. Borzumati, Y. Grossman, E. Nardi and Y. Nir,  
 Phys.Lett. {\bf B384} (1996) 123. 

\bibitem{Hempfling} R. Hempfling Nucl.Phys. {B478} (1996) 3.
\bibitem{white} B. de Carlos and P. White,  Phys. Rev. {\bf D54} (1996) 3427.
\bibitem{smirnov} A.Y. Smirnov and F. Vissani,  Nucl. Phys. {\bf B460} (1996) 37.
\bibitem{nilles} H.P. Nilles and N. Polonsky, Nucl. Phys. {\bf B499} (1997) 33.
\bibitem{kaplan} D.E. Kaplan and A.E. Nelson, hep-ph/9901254.
\bibitem{chunlee} E.J. Chun and J.S. Lee, hep-ph/9811201.
 \bibitem{Bednyakov} V. Bednyakov, A. Faessler and S. Kovalenko, 
 Phys. Lett. {\bf B442} (1998) 203.
 \bibitem{romao} J.C. Rom\~ ao, hep-ph/9907466.
  \bibitem{choi} S.Y. Choi, E.J. Chun, S.K. Kang and J.S. Lee, hep-ph/9903465.
\bibitem{drees} M. Dress, S. Pakvasa, X. Tata  and T. ter Veldhuis, 
Phys. Rev. {\bf D57} (1997) 5335.
\bibitem{bhattachar} S. Rakshit, G. Bhattacharyya and A. Raychadhuri, Phys. Rev. 
{\bf D59} 091701.
\bibitem{bhatta1} G. Bhattacharyya, H.V.
Klapdor-Kleingrothaus, H. Pas, hep-ph/9907432.

\bibitem{chunetal} E.J. Chun, S.K. Kang, C.W. Kim and J.S. Lee,  Nucl. Phys. 
{\bf B544} (1999) 89.
\bibitem{choihwang} K. Choi, E.J. Chun and K. Hwang, hep-ph/9811363.
\bibitem{Kong} O.C.W. Kong, Mod. Phys. Lett. {\bf A14} (1999) 903.

\bibitem{joshipura} A. Joshipura and S. Vempati, hep-ph/9903435.
A. Joshipura, V. Ravindran and S. Vempati, Phys. Lett {\bf B451} (1999) 98; 
A. Joshipura and M. Nowakowshi, Phys. Rev. {\bf D51} (1995) 2421.
\bibitem{datta} A. Datta, B. Mukhopadhyaya and S. Roy, hep-ph/9905549.
\bibitem{Kane} H. Haber and G. Kane,  Nucl. Phys. {\bf B272} (1986) 1.
\bibitem{davidson-king} S. Davidson and S. King, 
 Phys. Lett. {\bf B445} (1998) 191.

\bibitem{grossman2} Y. Grossman and H. Haber,  Phys. Rev. Lett. 
 {\bf 78} (1997) 3438; hep-ph/9906310.
\bibitem{abada-losada2} A. Abada and M. Losada, in preparation.

\bibitem{dreiner}H. Dreiner, hep-ph/9707435.
\bibitem{bhatta2} G. Bhattacharyya, Nucl. Phys. Proc. Suppl. {\bf 52A} (1997) 
83 and  hep-ph/9709395.
\bibitem{Chang-kee} Chang Kee Jung, Talk given at International Europhysics
 Conference on High-Energy Physics (EPS-HEP 99), Tampere, Finland, 15-21 Jul 1999. 
\bibitem{kongetal} M. Bisset, Otto C.W. Kong, C. Macenasu and Lynne Orr, hep-ph/9811498 (1998).

\bibitem{davidson} S. Davidson and J. Ellis,  Phys. Lett. {\bf B39} (1997) 210;
 Phys. Rev. {\bf D56} (1997) 4182. 





\end{thebibliography}
\end{document}